\documentclass[preprint2,dvipdfmx]{aastex}
\usepackage{here}
\usepackage{lscape}

\newcommand{\myemail}{seko@kusastro.kyoto-u.ac.jp}

\shorttitle{Molecular Gas in SF-galaxies at $z\sim1.4$}
\shortauthors{Seko et al.}

\begin{document}


\title{Properties of the Interstellar Medium in Star-Forming Galaxies at $z\sim1.4$ revealed with ALMA}


\author{Akifumi Seko\altaffilmark{1}, Kouji Ohta\altaffilmark{1}, Kiyoto Yabe\altaffilmark{2,3}, Bunyo Hatsukade\altaffilmark{2}, 
Masayuki Akiyama\altaffilmark{4}, Fumihide Iwamuro\altaffilmark{1}, Naoyuki Tamura\altaffilmark{3}, and Gavin Dalton\altaffilmark{5,6}}

\affil{$^{1}$Department of Astronomy, Kyoto University, Kitashirakawa-Oiwake-Cho, Sakyo-ku, Kyoto, 606-8502, Japan}
\email{\myemail}
\affil{$^{2}$National Astronomical Observatory of Japan, 2-21-1, Osawa, Mitaka, Tokyo, 181-8588, Japan}
\affil{$^{3}$Kavli Institute for the Physics and Mathematics of the Universe, Todai Institutes for Advanced Study, the University of Tokyo, Kashiwa, 277-8583, Japan (Kavli IPMU, WPI)}
\affil{$^{4}$Astronomical Institute, Tohoku University, 6-3 Aramaki, Aoba-ku, Sendai, 980-8578, Japan}
\affil{$^{5}$Department of Astrophysics, University of Oxford, Keble Road, Oxford OX1 3RH, UK}
\affil{$^{6}$STFC RALSpace, Harwell, Oxfordshire OX11 0QX, UK}

%
%

\begin{abstract}
We conducted observations of $^{12}$CO($J=5-4$) and dust thermal
continuum emission toward twenty star-forming galaxies on the main
sequence at $z\sim1.4$ using ALMA to investigate the properties of the
interstellar medium.  The sample galaxies are chosen to trace the
distributions of star-forming galaxies in diagrams of stellar
mass$-$star formation rate and stellar mass$-$metallicity.  We
detected CO emission lines from eleven galaxies.  The molecular gas mass
is derived by adopting a metallicity-dependent CO-to-H$_2$ conversion
factor and assuming a CO(5-4)/CO(1-0) luminosity ratio of 0.23.
Molecular gas masses and its fractions (molecular gas
mass$/$(molecular gas mass + stellar mass)) for the detected galaxies
are in the ranges of $(3.9-12)\times10^{10}~M_\odot$ and $0.25-0.94$,
respectively; these values are significantly larger than those in
local spiral galaxies.  The molecular gas mass fraction
decreases with increasing stellar mass; the relation holds for four
times lower stellar mass than that covered in previous studies, and
that the molecular gas mass fraction decreases with increasing
metallicity.  Stacking analyses also show the same trends.
The dust thermal emissions were clearly detected from 
two galaxies and marginally detected from five galaxies.  
Dust masses of the detected galaxies are $(3.9-38)\times10^{7}~M_\odot$.  
We derived gas-to-dust ratios and found they are 3-4 times larger 
than those in local galaxies.
The depletion times of molecular gas for the detected galaxies are
$(1.4-36)\times10^{8}~\mathrm{yr}$ while the results of the stacking
analysis show $\sim3\times10^{8}~\mathrm{yr}$.  
The depletion time tends to decrease with increasing stellar mass and 
metallicity though the trend is not so significant, 
which contrasts with the trends in local galaxies.
\end{abstract}


\section{Introduction}
The study of the properties of the interstellar medium (ISM) in
galaxies at high redshifts is indispensable for a complete
understanding of galaxy evolution.  Since galaxies evolve by
transforming gas into stars, revealing the amount of gas and its
fraction in a galaxy is important to trace galaxy evolution and to
understand a stage of the evolution.  Since stars eject metals through
supernovae explosions and/or mass loss, the dust mass is considered to
reflect the star-formation history in a galaxy.  In addition, dust
plays an important role in formation of hydrogen molecules and in the
cooling of the ISM.  Thus, revealing the dust mass and gas-to-dust
ratios is also important to understand galaxy evolution.

Most star-forming galaxies form a sequence in the stellar
mass$-$star-formation rate (SFR) diagram at each redshift at least up
to $z\sim2.5$ \citep[e.g.,][]{Noes07, Dadd07, Rodi10, Whit12, Whit14},
which is called the ``main sequence" of star-forming galaxies.  It is
particularly important to unveil the nature of the ISM in main
sequence galaxies at $z=1-2$, because this epoch is the peak of the
cosmological evolution of the star-formation rate density
\citep[e.g.,][]{Hopk06, Mada14}.

Recently, high sensitivity radio telescopes have enabled us to detect
CO emission from massive ($M_\ast>2.5\times10^{10}~M_\odot$ assuming a
Chabrier initial mass function \citep[IMF:][]{Chab03}) star-forming
galaxies at $z=1-2.5$ \citep[e.g.,][]{Dadd08, Dadd10a, Dadd15, Tacc10,
  Tacc13, Genz12, Genz13}.  \citet{Tacc10, Tacc13} derived the
molecular gas masses in star-forming galaxies at $z=1-2.5$
($M_\mathrm{mol}\sim10^{10-11}~M_\odot$) adopting the Galactic
CO-to-H$_2$ conversion factor
($\alpha_\mathrm{CO}=4.36~M_\odot~(\mathrm{K~km~s^{-1}~pc^2})^{-1}$
including helium mass).  They found the molecular gas masses to be larger
than those in the present-day massive spiral galaxies
\citep[$M_\mathrm{mol}\sim10^{8.5-10}~M_\odot$; e.g.,][]{Sain11a}, and
that the molecular gas mass fractions [$f_\mathrm{mol} =
  M_\mathrm{mol}/(M_\mathrm{mol} + M_\mathrm{star})$] in this stellar
mass range are $\sim30-50\%$, after correcting for a bias to larger
specific SFR. This range is significantly higher than the typical
value in the local massive spiral galaxies ($f_\mathrm{mol}\sim8\%$).
They also found molecular gas mass fraction tends to decrease with
increasing stellar mass.

The recent advent of {\it Spitzer}/MIPS and {\it Herschel}/PACS, SPIRE
enables us to investigate dust emission in the mid- and far-infrared
from high redshift galaxies on the main sequence up to $z\sim2$
\citep[e.g.,][]{Elba11}.  \citet{Magd12a, Magd12b} derived dust masses
in main sequence galaxies at $z=1-3$
($M_\mathrm{dust}\sim10^{8-9}~M_\odot$) and showed that the dust mass
is larger than that found in present-day spiral galaxies
\citep[$M_\mathrm{dust}\sim10^{6.5-8}~M_\odot$; e.g.,][]{Remy14}.  To
investigate the molecular gas content at this redshift, the molecular
gas mass was estimated from the dust mass by assuming the gas-to-dust
ratio \citep{Magd11, Magd12a, Magd12b, Magn12}.  These studies also
found that galaxies at this redshift show larger molecular gas mass
fractions and that the fraction decreases with increasing stellar
mass.

Another key parameter that can be used to trace galaxy evolution is
the gas-phase metallicity because it reflects past star-forming
activity.  The relation between stellar mass and metallicity
(hereafter mass-metallicity relation) has been studied at $z=0-3$;
galaxies with larger stellar mass tend to have higher metallicity
\citep{Trem04, Erb06, Maio08, Mous11, Zahi11, Zahi13, Yabe12, Yabe14}.
The mass-metallicity relation seems to be smoothly evolved from
$z\sim3$ to $z\sim0$; galaxies at higher redshift have systematically
lower metallicity at a fixed stellar mass \citep[e.g.,][]{Zahi13},
although the trend may be explained without invoking evolution in
terms of the fundamental metallicity relation \citep[e.g.,][]{Mann10,
  Lara10}.  The molecular gas mass fraction is expected to be related
to metallicity.  At high redshift, however, the relation is unknown,
because there are very few CO observations toward the main sequence
galaxies with known metallicities.  In the local universe, the
gas-to-dust ratio also shows a dependence on gas-phase metallicity;
the ratio is $\sim150$ at solar metallicity and increases with
decreasing metallicity \citep[e.g.,][]{Lero11, Remy14}.  In galaxies
at $z=1-3$, the ratio is comparable to or a factor of 2 larger than
that in nearby galaxies at a fixed metallicity \citep{Sain13, Seko14}.
However, the number of galaxies with known metallicities at high
redshifts that have measurements of both CO and dust remains very
limited.

\begin{figure*}[tb]
\begin{center}
	\epsscale{2.0}
		\plotone{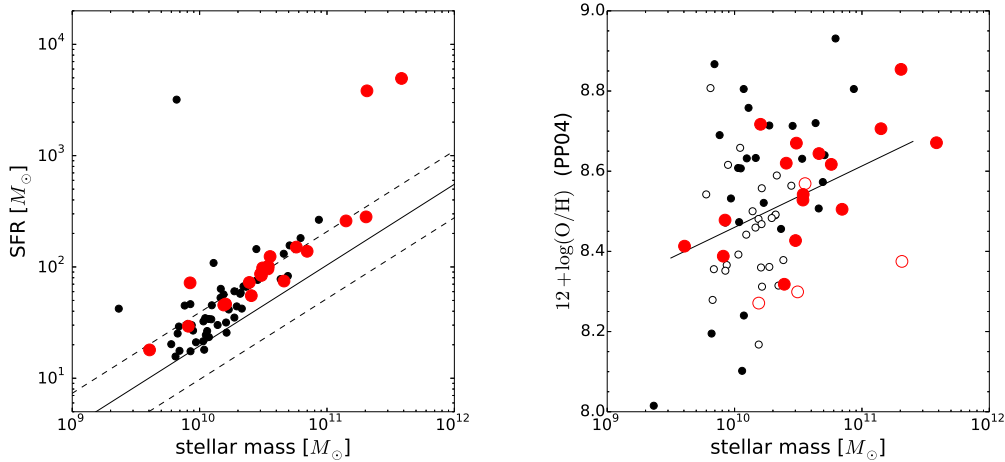}
		\caption{(Left) Sample galaxies in the stellar
                  mass-SFR diagram.  Circles show 71 H$\alpha$
                  emitting galaxies at $z\sim1.4$ by \citet{Yabe12}
                  (see text for details).  Among them, large (red)
                  circles show 20 sample galaxies observed with ALMA.
                  SFRs are derived from extinction corrected UV
                  luminosity densities.  
                  Solid line represents the main sequence at $z\sim1.4$ 
                  by \citet{Spea14} after correcting for the difference in IMFs used. 
                  Dashed lines show the scatter.
                  (Right) The same sample
                  galaxies in the stellar mass-metallicity diagram.
                  The metallicities are derived from H$\alpha$ and
                  [NII] emission lines \citep[][PP04
                    calibration]{PP04}.  Large (red) symbols show the
                  ALMA sample galaxies.  Objects with [NII]
                  $\lambda6584$ lines with $\mathrm{S/N}>1.5$ and
                  $\mathrm{S/N}<1.5$ are indicated by filled and open
                  circles, respectively, i.e., open circles represent
                  upper limits on metallicity. 
                  Solid line represents the mass$-$metallicity relation at $z\sim1.4$
                  by \citet{Yabe14}.}
\label{fig: sample}
\end{center}
\end{figure*}

The high sensitivity of the Atacama Large Millimeter/submillimeter
Array (ALMA) enables us to study the ISM in star-forming galaxies at
high redshift.  In this study, we aim to reveal the relations between
the molecular gas mass or its fraction and the stellar mass or
metallicity for main sequence galaxies at $z\sim1.4$.  In addition, we
also examine the dust properties and the gas-to-dust ratio at this
redshift.  The uniqueness of our sample is that the metallicity has
already been measured for each galaxy with near-infrared spectroscopy as
described in section \ref{sec: sample}.  Thus, we can also adopt a
metallicity-dependent CO-to-H$_2$ conversion factor.  In section
\ref{sec: observation & reduction}, the observations and data reduction
are described, with source detection for CO emission lines and dust
thermal emissions presented in section \ref{sec: source
  detection}.  Results for CO emissions and dust thermal emissions are
shown in sections \ref{sec: result CO} and \ref{sec: result
  dust}, respectively.  Results for gas-to-dust ratio are given in
section \ref{sec: GDR}.  Throughout this paper, we adopt the standard
$\Lambda$-CDM cosmology with $H_{0} = 70~\mathrm{km~s^{-1}~Mpc^{-1}}$,
$\Omega_{M} = 0.3$, and $\Omega_{\Lambda}=0.7$.

\section{Samples}\label{sec: sample}
Sample galaxies used in this study are taken from \cite{Yabe12}. 
\citet{Yabe12} used far-UV and near-UV data 
taken from Galaxy Evolution Explorer (GALEX) archived image (GR6), 
U-band images taken from the Canada-France-Hawaii Telescope Legacy Survey (CFHTLS) wide surveys,
optical images ($B$, $V$, $R_{C}$, $i^{'}$, and $z^{'}$-band) 
from the Subaru-XMM/Newton Deep Survey (SXDS), 
the near-infrared images ($J$, $H$, and $K_{s}$-band) 
from the DR8 version of the UKIRT Infrared Deep Sky Survey (UKIDSS) 
Ultra Deep Survey (UDS), 
and {\it Spitzer} Infrared Array Camera (IRAC) images 
(3.6~$\mu\mathrm{m}$, 4.5~$\mu\mathrm{m}$, 5.8~$\mu\mathrm{m}$, and 8.0~$\mu\mathrm{m}$)
from the {\it Spitzer} public legacy survey of the UKIDSS UDS (SpUDS). 
Photometric redshift was derived with Hyperz \citep{Bolz00}. 
Stellar mass was derived by spectral energy distribution (SED) fitting \citep{Sawi12} 
with the optical to mid-infrared data 
by employing the population synthesis model by \citet{BC03}. 
SFRs are derived from rest-frame UV luminosity densities, 
corrected for the dust extinction estimated from the rest-frame UV-slope.  
In both stellar mass and SFR, the Salpeter IMF \citep{Salp55} 
with a mass range of $0.1-100~M_\odot$ was adopted. 
\citet{Yabe12} selected star-forming galaxies at $z\sim1.4$ with
$K_\mathrm{s}<23.9~\mathrm{mag}$ (AB magnitude) and stellar mass
$\geq10^{9.5}~M_\odot$. 

Using this sample, \citet{Yabe12} made near-IR spectroscopic
observations with the Fiber Multi-Object Spectrograph (FMOS;
\citealt{Kimu10}) on the Subaru telescope.  H$\alpha$ emission lines
were significantly detected for 71 star-forming galaxies.  
The stellar mass and SFR for these are shown in Figure \ref{fig: sample}. 
The galaxies locate on the main sequence of star-forming galaxies 
at $z\sim1.4$ by \citet{Spea14}. 
However, the galaxies might be biased to slightly larger SFR, 
because \citet{Yabe12} selected galaxies with expected H$\alpha$ flux 
larger than $1.0\times10^{-16}~\mathrm{erg~s^{-1}~cm^{-2}}$.
The gas metallicities are derived from the N2 method by using 
the H$\alpha$ and [NII]$\lambda$ 6584 emission lines \citep{PP04}.  
It should be noted that \citet{Yabe15} showed that the nitrogen-to-oxygen
abundance ratio in star-forming galaxies at this redshift is
significantly higher than the local value at a fixed metallicity and
stellar mass, and thus there is a possibility that the metallicity
derived with N2 method is systematically overestimated by
$0.1-0.2$~dex.

We selected 20 of these 71 galaxies as ALMA targets to cover a wide
range of stellar mass ($4\times10^9$ - $4\times10^{11}~M_\odot$) and
metallicity ($12+\log(\mathrm{O/H}) = 8.2-8.9$) and to trace these
distributions rather uniformly.  The selected sample galaxies are
shown with large (red) circles in Figure \ref{fig: sample} and listed
in Table \ref{table: sample}.  Almost all of our sample galaxies
lie on the main-sequence.  The two most massive galaxies (SXDS1\_35572
and SXDS1\_79307) show very high SFR, and these may not be on the main
sequence.  The SFR estimation may not be correct for these two
galaxies due to, for example, the uncertainty in the dust extinction.
Nevertheless, we included these two galaxies to examine their nature.~\footnote[1]{It was turned out 
later that no far-infrared continuum toward these galaxies was detected by {\it Herschel}, 
i.e., $\mathrm{SFR} \leq 50~M_\odot~\mathrm{yr^{-1}}$ \citep{Elba11}. 
SFR of SXDS1\_35572 derived from the H$\alpha$ luminosity is 
$16~M_\odot~\mathrm{yr^{-1}}$ (extinction uncorrected) 
and $537~M_\odot~\mathrm{yr^{-1}}$ (extinction corrected) 
which is larger than the upper limit from {\it Herschel} data. 
Thus, the correction of dust extinction may be overestimated. 
SFR of SXDS1\_79307 from the extinction uncorrected H$\alpha$
luminosity is $427~M_\odot~\mathrm{yr^{-1}}$. 
A reason for this large SFR may be that an OH airglow sky emission comes 
very close to the position of the H$\alpha$ 
and we could not completely remove the sky emission. 
}  
Because the X-ray luminosities ($L_\mathrm{X(2-10~keV)}$) of our sample galaxies are
less than $10^{43}~\mathrm{erg~s^{-1}}$, they are not X-ray bright
active galactic nuclei.


\section{Observations and data reduction} \label{sec: observation & reduction}
\subsection{Observations}
We made $^{12}$CO($J=5-4$) observations toward the 20 galaxies using
ALMA.  The observations were carried out in 2012 August 9, 11, 15, and
26 during the ALMA cycle0 session (ID=2011.0.00648.S, PI=K. Ohta).
The on-source time for each galaxy was $8-15~\mathrm{min}$.  The
number of 12~m antennas was 23-25.  The observed frequencies were
222.094~GHz to 252.583~GHz (band-6).  We used four correlator setups.
The frequencies of local oscillator in each setup were 231.198~GHz,
236.168~GHz, 240.380~GHz, and 244.166~GHz.  To cover the CO emission
lines of all the sample galaxies, we set three or four spectral
windows (SPWs) in each correlator setup.  Each SPW had a bandwidth of
1.875~GHz.  The spectral resolution was 488.28~kHz, corresponding to a
velocity resolution of $0.58 - 0.66~\mathrm{km~s^{-1}}$ at the
observed frequency range.  The FWHM of primary beam was about 26$''$.
The flux calibration was made with the Ganymede, Uranus, and Callisto.
The phase calibrator was J$0204-170$.  The bandpass calibrator was
J$2253+161$.

\subsection{Data reduction}
Data reduction was carried out with the Common Astronomy Software
Applications \citep[CASA:][]{McMu07} version 4.2 package in a standard
manner.  The delivered data which were calibrated had problems; the
coordinates of the phase calibrator were wrong for three correlator
setups (15 target galaxies) and the data of the flux calibrator for a
SPW in one correlator setup was flagged for some unknown reason.  The
coordinates of the phase calibrator were found to be systematically
shifted by the $0''.3$ in right ascension and $0''.04$ in declination.
Since this shift is not negligible for stacking analysis, we made
re-calibrations with the corrected coordinates of the phase calibrator
and with the interpolated value for the flagged data of the flux
calibrator.  We used the 2012 models of the Solar system objects for
flux calibrations.

We subtracted the continuum emission in uv-data by using the CASA task
UVCONTSUB.  Continuum maps were made by combining both the lower side
band (LSB) and upper side band (USB) data in line-free frequencies.
The maps were made with the CASA task CLEAN with natural weighting.
The center position of each map coincided with the centroid of the
galaxy in the K-band image.  The number of iteration was zero (i.e.,
dirty maps) because the signal-to-noise ratios of the detected sources
were not high and we were afraid the uncertainty in the maps increases
by CLEAN.  In addition, since we will discuss the results both for
individual galaxy and stacked data below, we employ the same reduction
procedure for detected and non-detected galaxies.  It should be also
noted that no strong sources exist in the fields of view.  The
synthesized beam size was $0''.6-1''.3$ (Table \ref{table: CO-2}).  The
noise level of continuum map was $0.04-0.1~\mathrm{mJy~beam^{-1}}$
(Table \ref{table: CO-2}).  For the CO emission studies, we made channel
maps from the continuum subtracted data with a velocity resolution of
$\sim50~\mathrm{km~s^{-1}}$; the noise level of each channel map was
$0.5-1~\mathrm{mJy~beam^{-1}}$.

\begin{figure*}
	\epsscale{2.0}
		\plotone{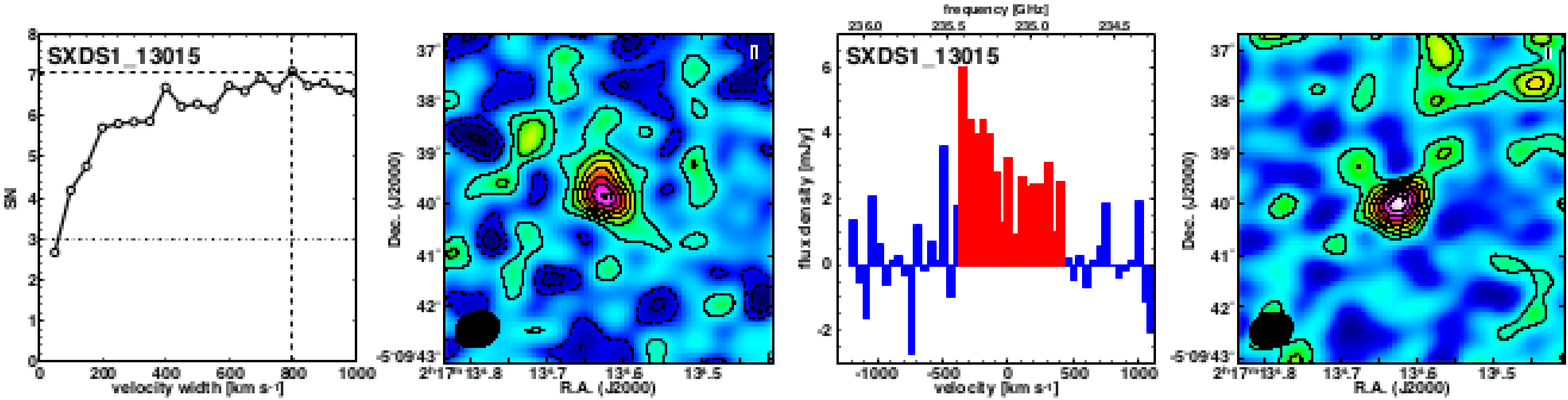}
		\plotone{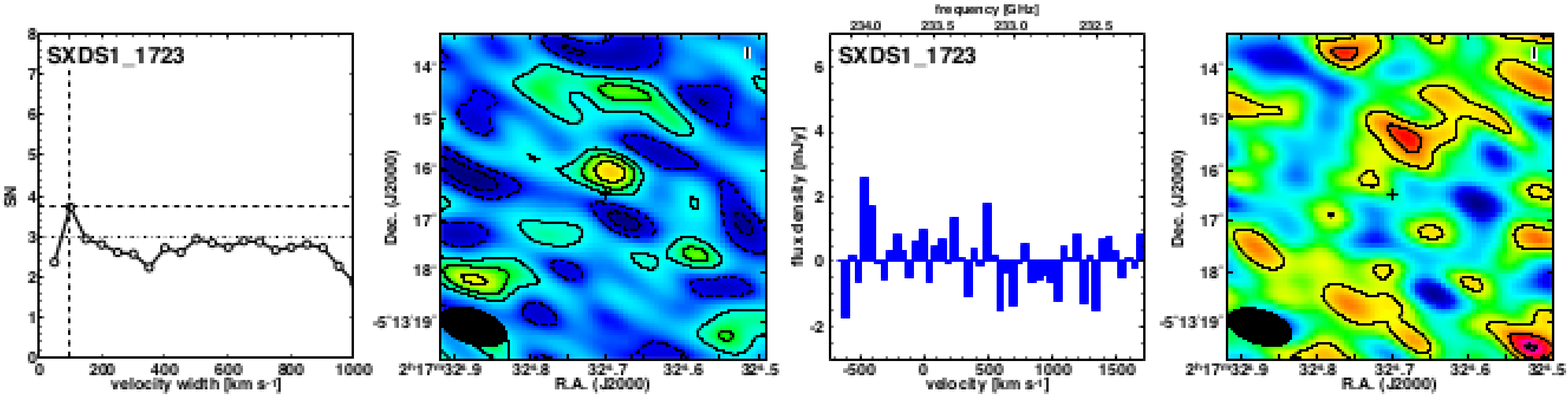}
		\plotone{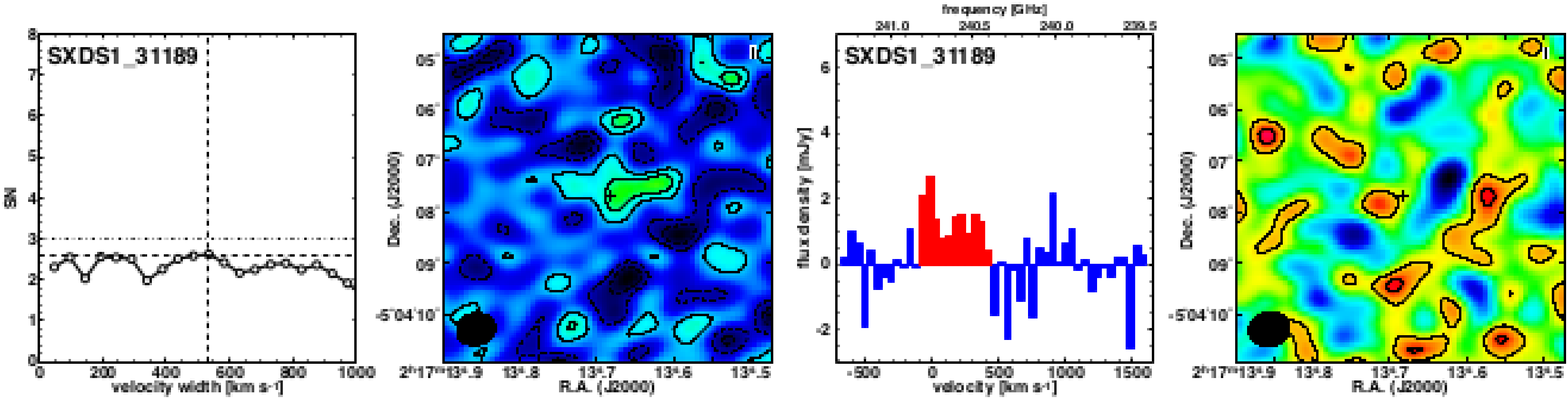}
		\plotone{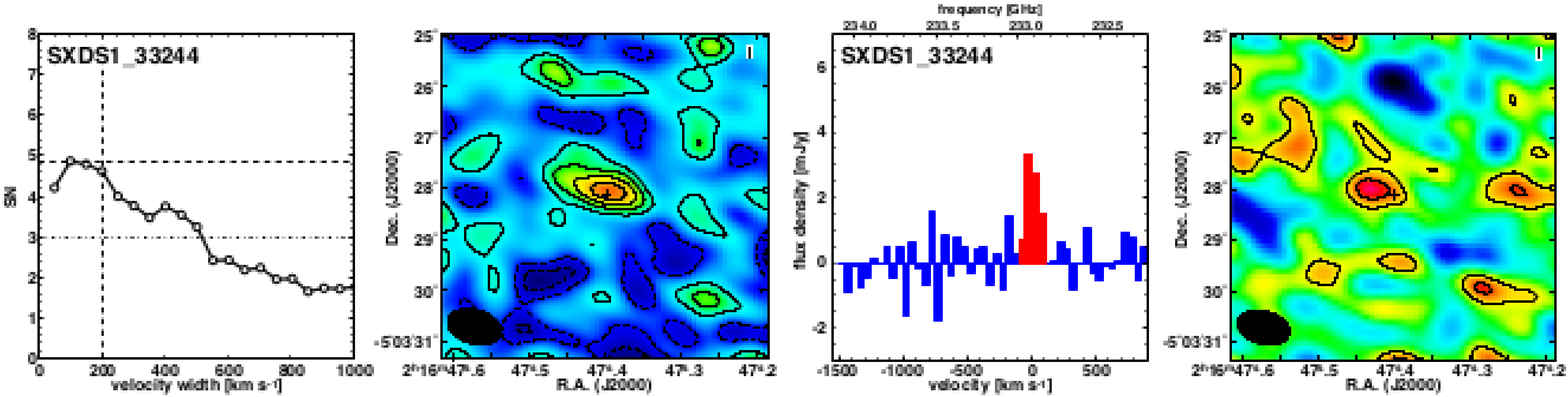}
		\caption{(Left) Growth curve of signal-to-noise ratio
                  against integrated velocity width.  Horizontal
                  dashed line and dashed-dotted line refer to the peak
                  SN and the SN of 3, respectively.  Vertical dashed
                  line refers to the integrated velocity width
                  adopted.  (Center Left) Integrated CO(5-4) intensity
                  map.  The integrated velocity width is shown with a
                  vertical dashed line in the growth curve of
                  signal-to-noise ratio.  Contours represent
                  $-2\sigma$, $-1\sigma$ (dashed lines), $1\sigma$,
                  $2\sigma$, $3\sigma$, \ldots (solid lines).  The
                  cross refers to the peak position in the K-band
                  image. The filled black ellipse at a bottom left
                  corner shows the synthesized beam size.  (Center
                  Right) CO(5-4) spectrum.  The zero velocity is
                  derived from the spectroscopic redshift of the
                  H$\alpha$ line.  For the CO detected galaxies, the
                  spectra are made in the region where SN is larger
                  than 1 around the source, and the CO emission line
                  is shown with red color.  For non-detected galaxies,
                  the spectra are made in the central box.  (Right)
                  Continuum map. Contours represent $1\sigma$,
                  $2\sigma$, $3\sigma$, \ldots (solid lines).  The
                  cross refers to the peak position in the K-band
                  image. The filled black ellipse again shows the
                  synthesized beam size. }
\label{fig: map spectrum sn-plot}
\end{figure*}

\begin{figure*}
\begin{center}
	\epsscale{2.0}
		\plotone{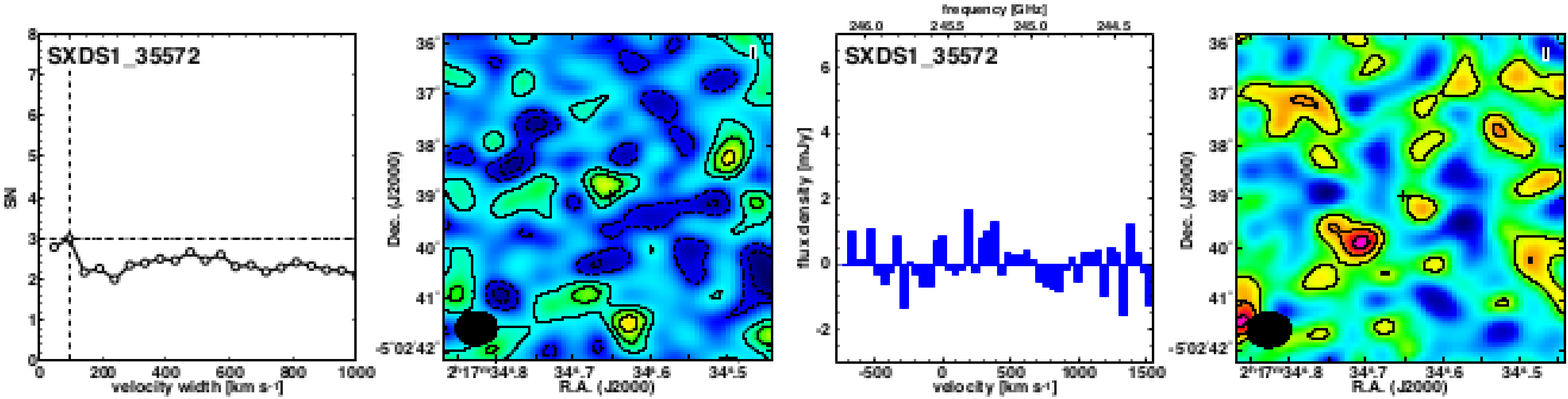}
		\plotone{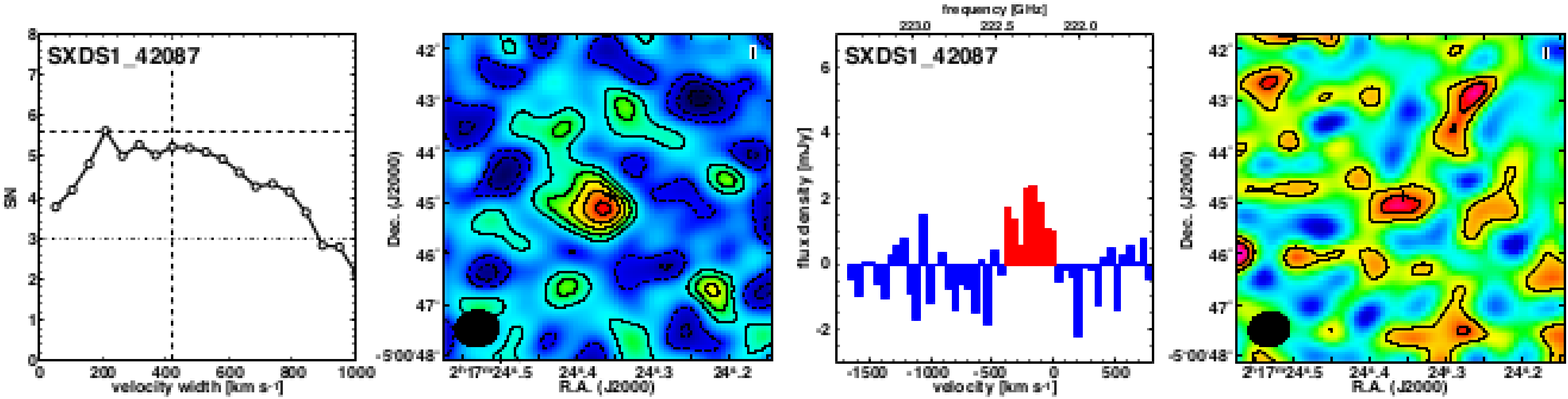}
		\plotone{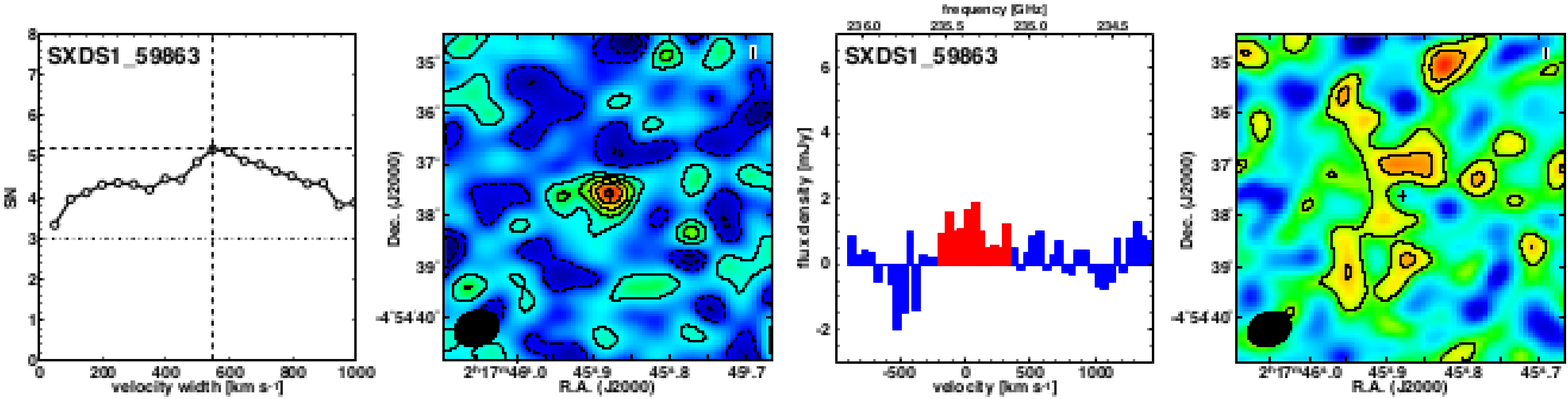}
		\plotone{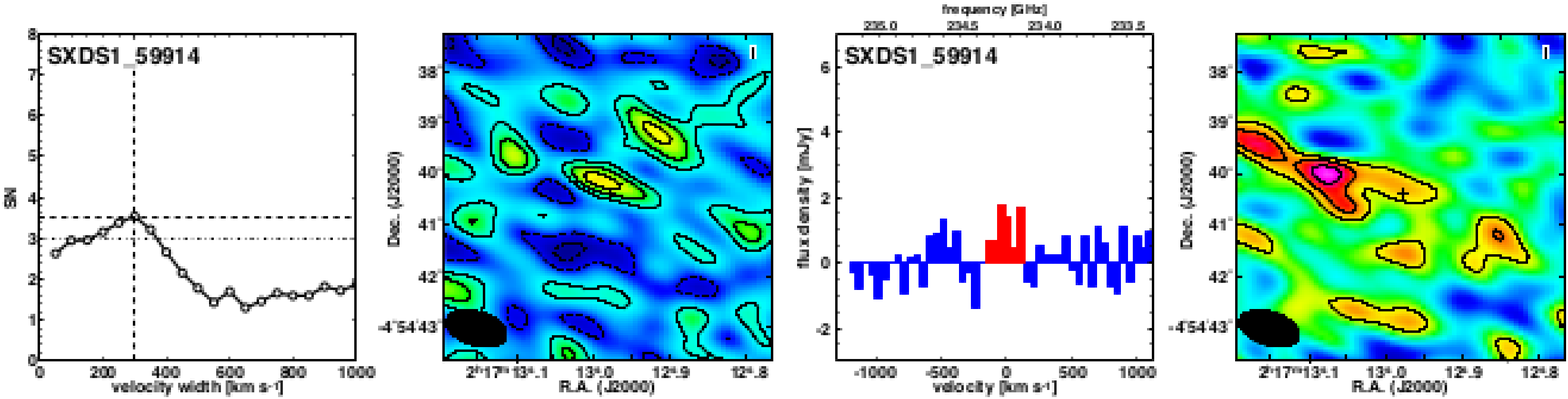}
		\plotone{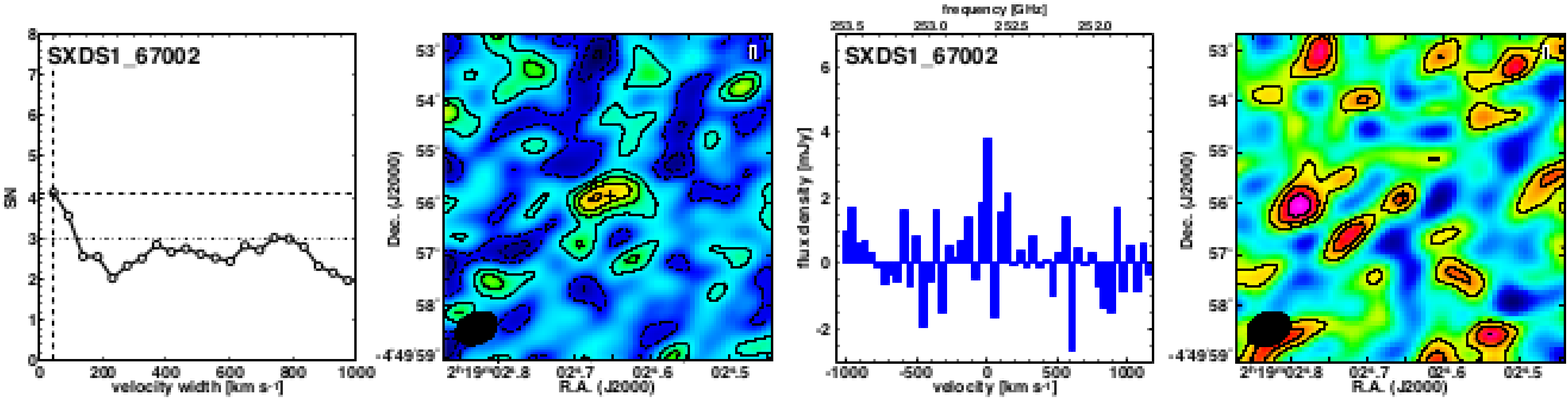}
		\caption{Same as Figure \ref{fig: map spectrum sn-plot} but for other galaxies. }
\label{fig: appendix}
\end{center}
\end{figure*}

\begin{figure*}
\begin{center}
	\epsscale{2.0}
		\plotone{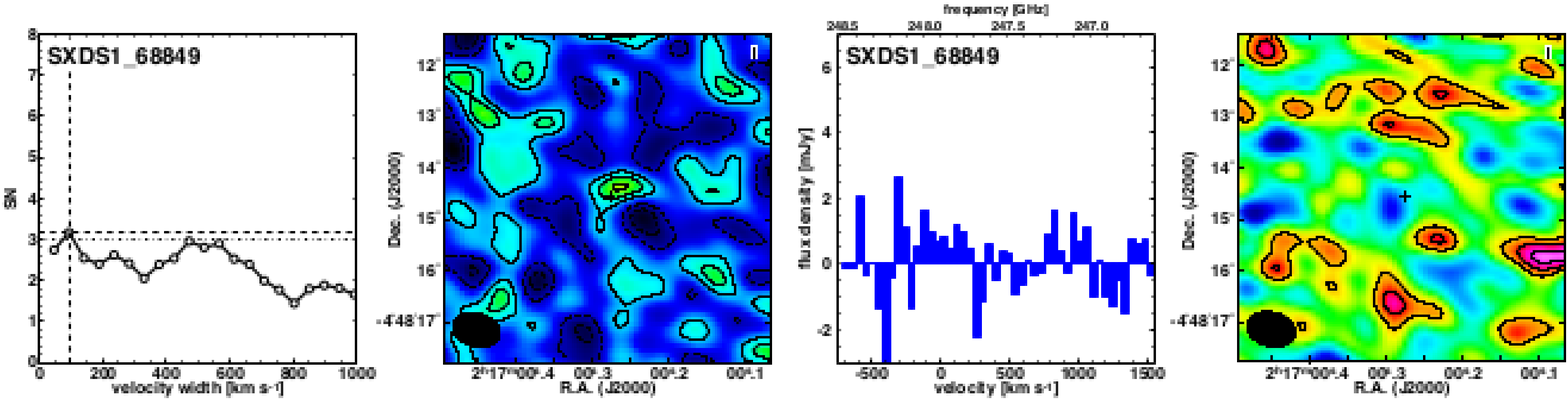}
		\plotone{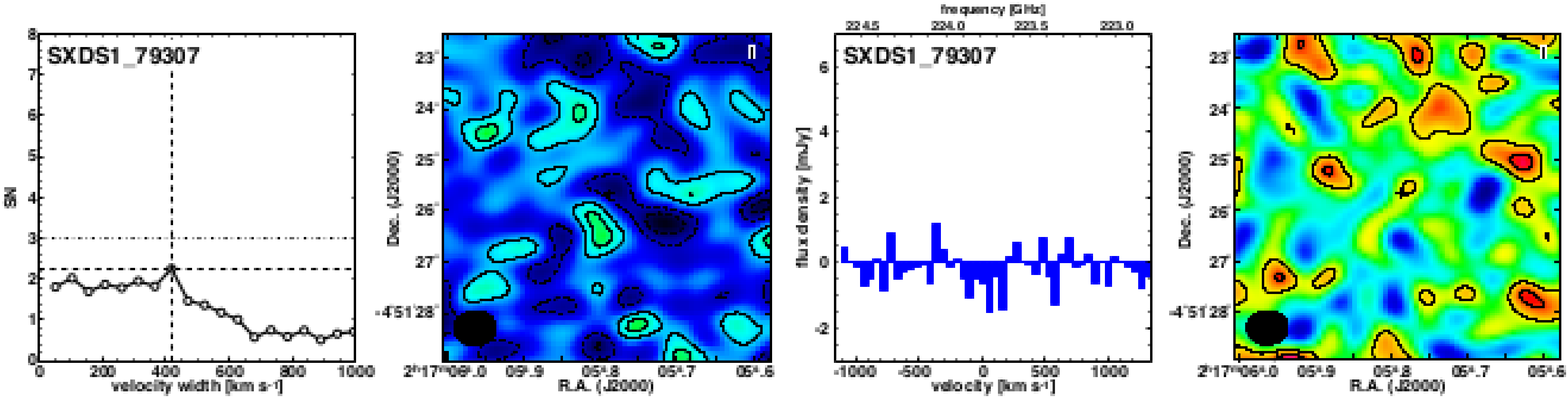}
		\plotone{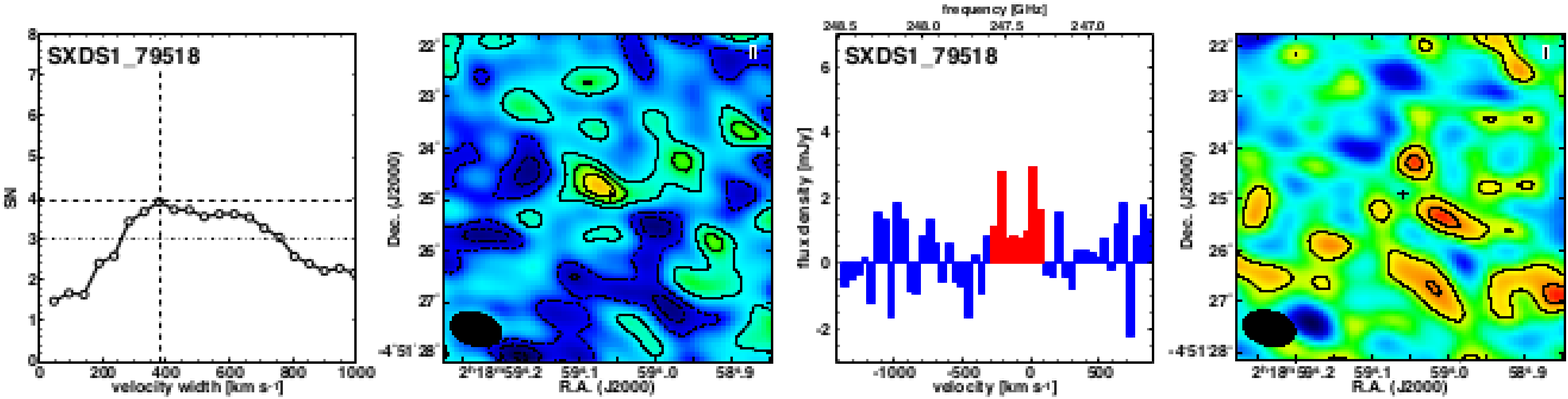}
		\plotone{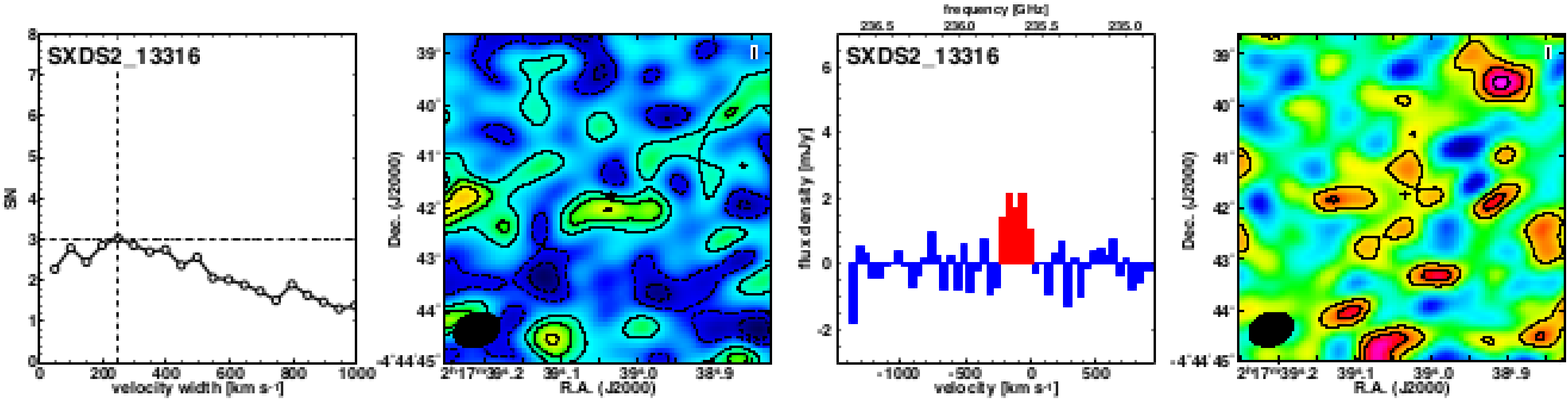}
		\plotone{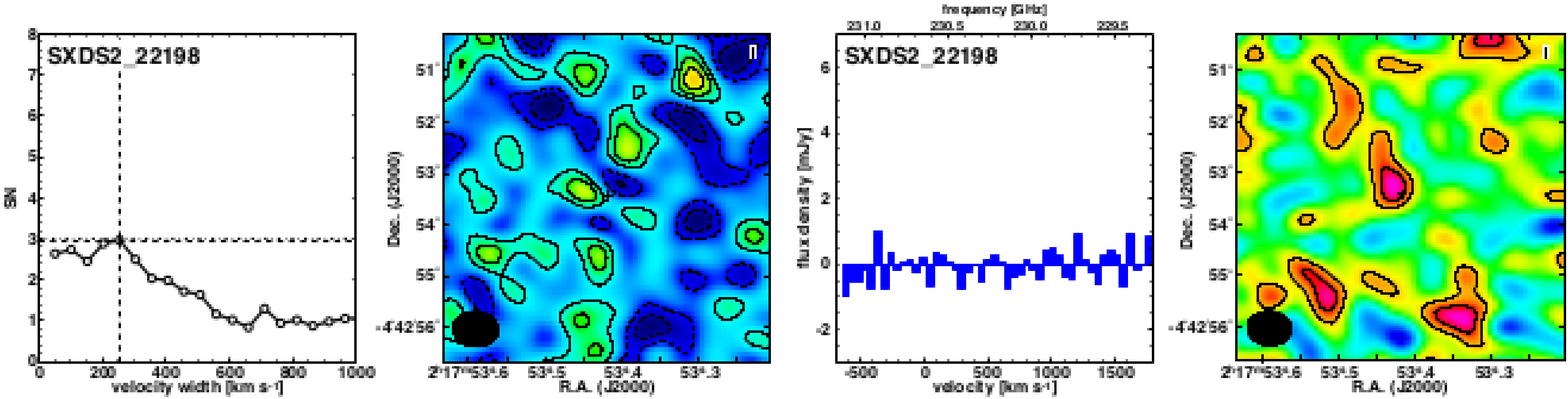}
		\caption{Same as Figure \ref{fig: map spectrum sn-plot} but for other galaxies. (continued)}
\label{fig: appendix}
\end{center}
\end{figure*}

\begin{figure*}
\begin{center}
	\epsscale{2.0}
		\plotone{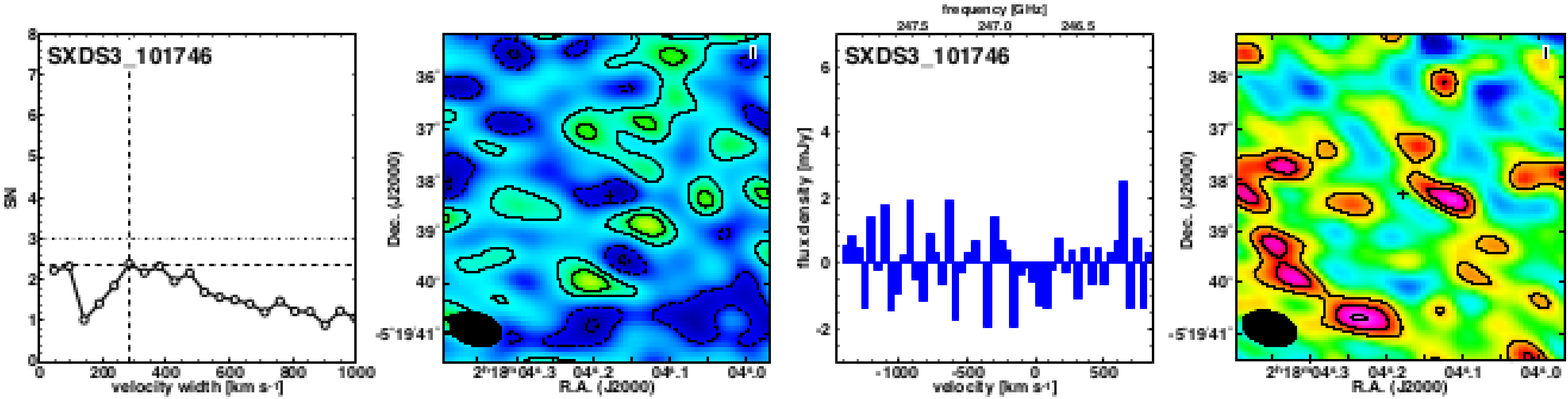}
		\plotone{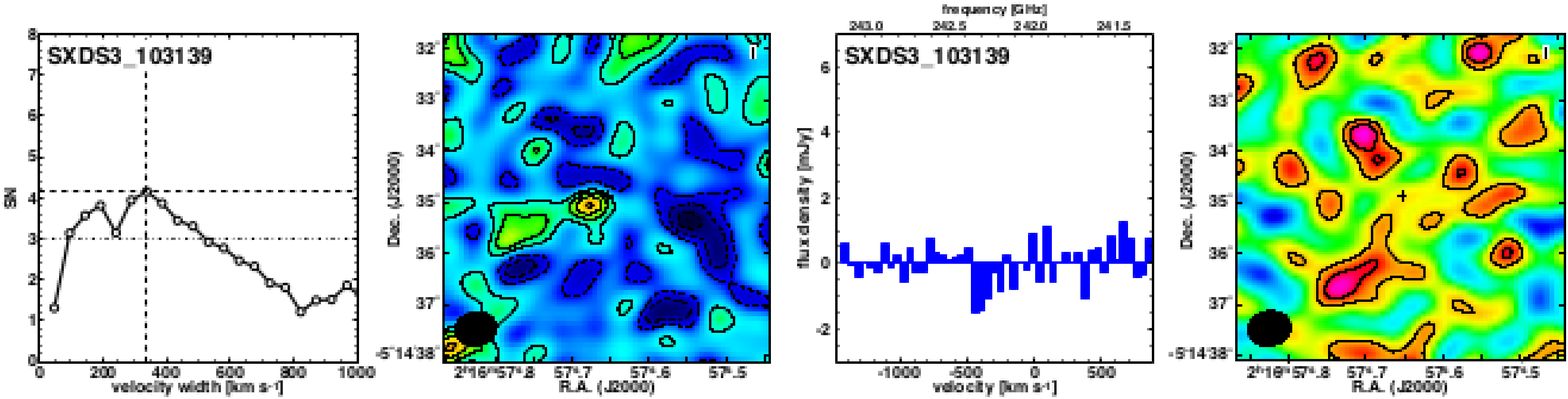}
		\plotone{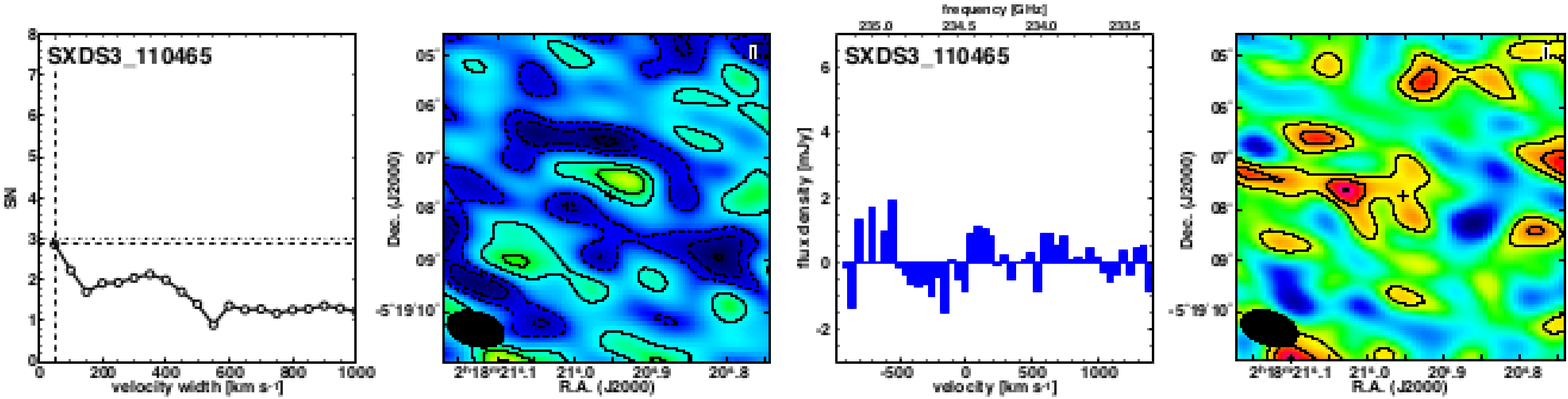}
		\plotone{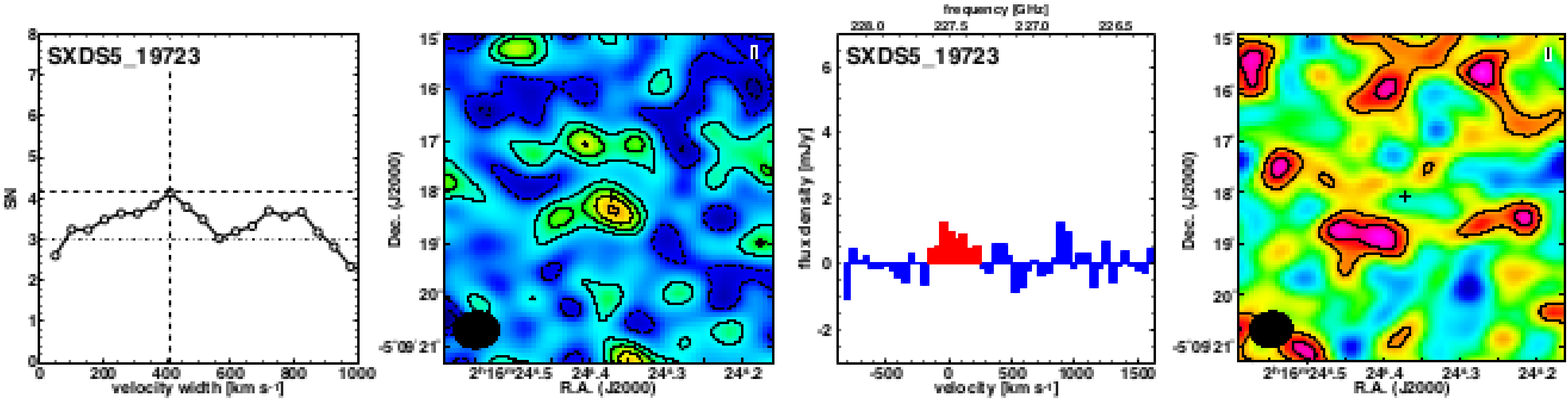}
		\plotone{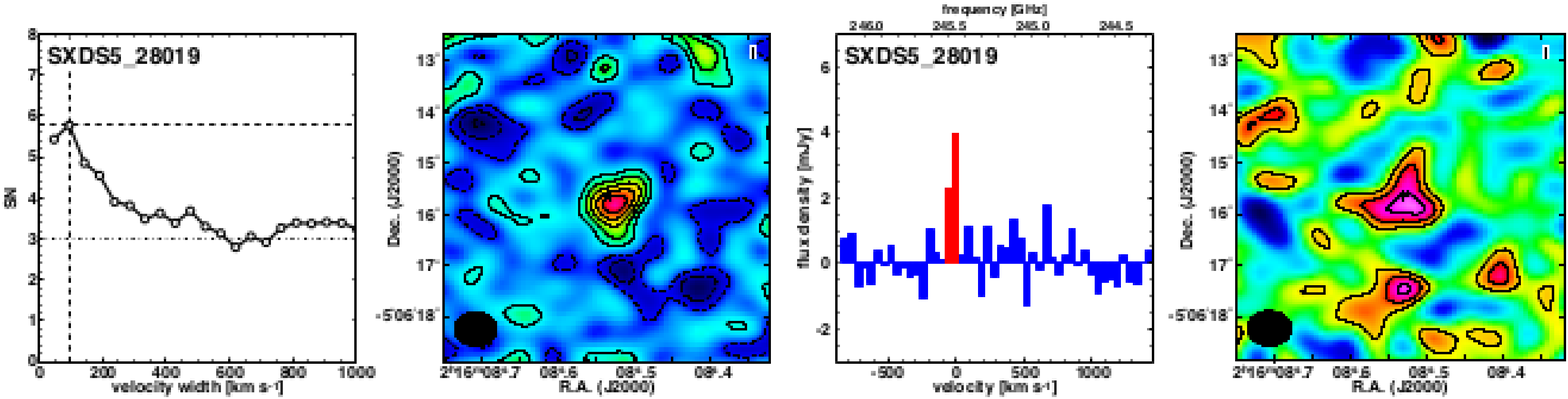}
		\plotone{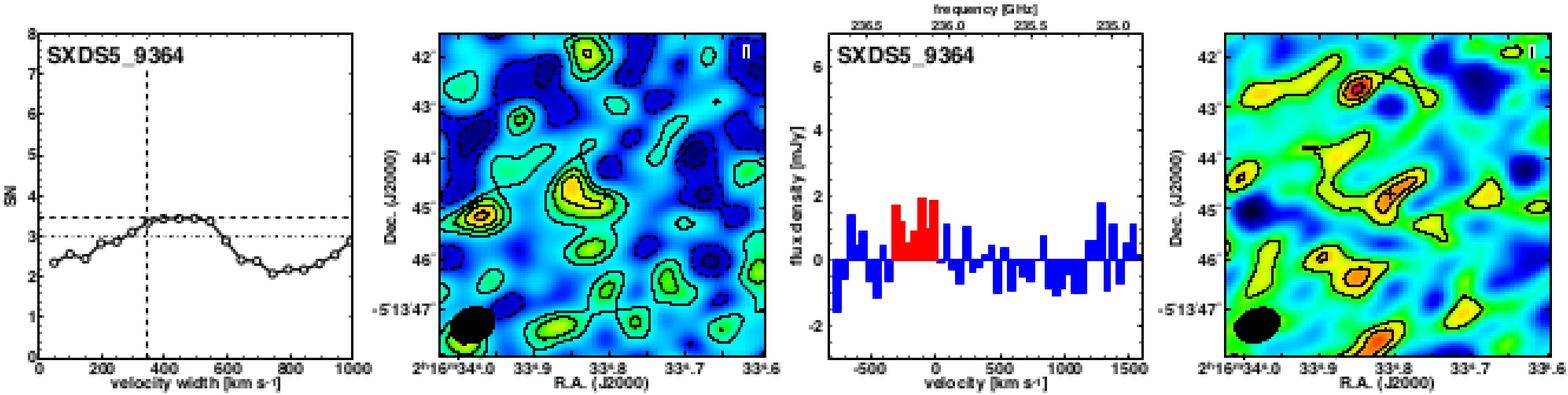}
		\caption{Same as Figure \ref{fig: map spectrum sn-plot} but for other galaxies. (continued)}
\label{fig: appendix}
\end{center}
\end{figure*}

\section{Source detection} \label{sec: source detection}
The following procedure is employed to search for CO emission lines:
The redshifts of our sample galaxies are known from the near-IR
spectroscopic observations of H$\alpha$ emission lines.  Thus we
derive the ``zero velocity" by referring this redshift.  We define the
``central box", which is the region within $\pm0''.5$ in right
ascension and declination from the map center.  Since the angular
resolution of our observations is $0''.6-1''.3$, the peak position of any
CO emission is reasonably expected to be in the central box if
signal-to-noise ratio (SN) is $\geq2$.

Using the CASA task IMMOMENTS, we make zeroth-order moment maps
(integrated intensity maps) by changing the velocity width ($\Delta v =
50-1000~\mathrm{km~s^{-1}}$) centered at the zero velocity.  We
measure the SN in the central box for each integrated intensity map and
make a SN growth curve against the velocity width.  We regard the
feature as candidate for detection if the following criteria are
satisfied: the highest SN in the growth curve is $>3$, the
SNs within a velocity width $\pm50~\mathrm{km~s^{-1}}$ from the
velocity width giving the highest SN are also $>3$, and the
peak positions in the maps lie within the central box.

Since the accuracy of the zero velocity is
$\sim150~\mathrm{km~s^{-1}}$, there may be a more appropriate choice
of zero velocity.  Thus, we change the ``central velocity" of the
integrated intensity maps within $\pm200~\mathrm{km~s^{-1}}$ from the
zero velocity and repeat the analysis for all targets.

Among the two analyses mentioned in the above two paragraphs, we take
the combination of the central velocity and velocity width which
produces the highest-SN peak in the central box.  Ten such sources are
regarded as candidates for detection.

Next, we make a spectrum of the candidate in the region with SN $>1$
around the candidate in the integrated intensity map.  If CO emission
is also seen adjacent to the velocity width adopted above, we made the
integrated intensity map including this velocity range and check
whether the map satisfies the criteria mentioned above.  Finally, when
SN in the spectrum smoothed with the integrated velocity width is
$>3$, we consider the CO emission to be detected.

In the case of SXDS1\_31189, the SNs in the integrated intensity maps
for all velocity widths are slightly less than 3, but the emission
line is clearly seen in the spectrum at zero velocity.  Thus, we
consider the CO emission line to be detected.  In the case of
SXDS2\_13316, the SN is $\sim3$ at the velocity width of
$250~\mathrm{km~s^{-1}}$, but SN is slightly less than 3 when the
velocity width changes by $\pm50~\mathrm{km~s^{-1}}$.  Again, the
emission line is very clearly seen in the spectrum.  Thus, we
also consider this CO emission line to be detected.

In Figure \ref{fig: map spectrum sn-plot}, we show the SN growth
curves, the integrated intensity maps made with the velocity width
shown in the growth curves, and the CO($J=5-4$) line profiles.  The
peak positions of all the detected source are within $\pm0''.3$ in
right ascension and declination from the map center, confirming that
the size of the central box is appropriate.

For the detection of dust thermal emissions, we use the continuum map
(right panels of Figure \ref{fig: map spectrum sn-plot}).  We consider
detections for cases where the peak SN in the central box is $>3$, and
marginal detections if the peak SN is $2-3$.

\begin{figure*}
\begin{center}
	\epsscale{2.0}
		\plotone{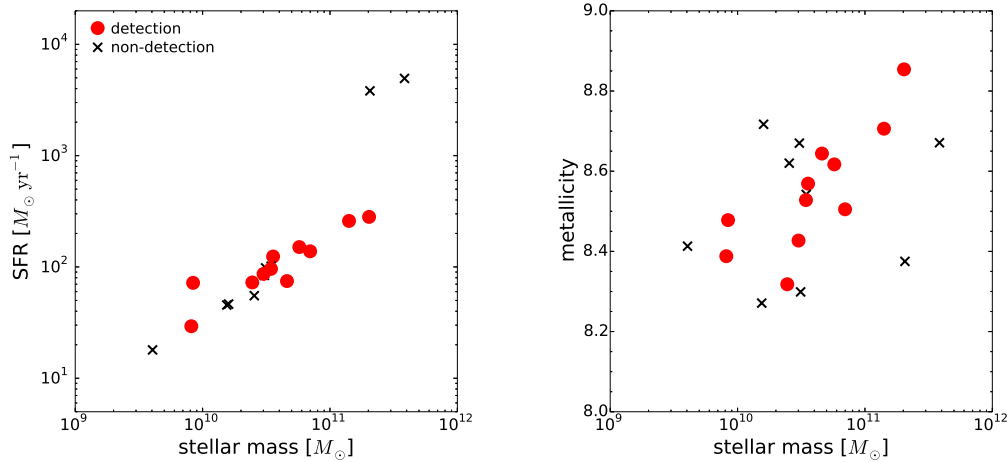}
		\caption{The location of the observed galaxies in the
                  stellar mass$-$SFR diagram (Left) and in the stellar
                  mass$-$metallicity diagram (Right).  Filled (red)
                  circles refer to the galaxies for which CO($J=5-4$)
                  emission lines are detected.  Crosses show the
                  non-detections. }
\label{fig: detected sources}
\end{center}
\end{figure*}

\section{Results for CO emission lines} \label{sec: result CO}
\subsection{Individual galaxy}
We detected CO($J=5-4$) emission lines from 11 galaxies.  These
galaxies are shown with filled (red) circles in the stellar mass$-$SFR
and stellar mass$-$metallicity diagrams (Figure \ref{fig: detected
  sources}).  The CO lines tend to be detected for galaxies with more
massive/higher SFR on an average.  No clear dependence on metallicity
is seen, though the average metallicity of the detected galaxies is
slightly larger than that of the non-detected galaxies.  For the most
massive two galaxies (SXDS1\_35572 and SXDS1\_79307), CO emission
lines were not detected, suggesting that the estimation of SFR is not
correct for these two galaxies.

\begin{figure*}
\begin{center}
	\epsscale{2.0}
		\plotone{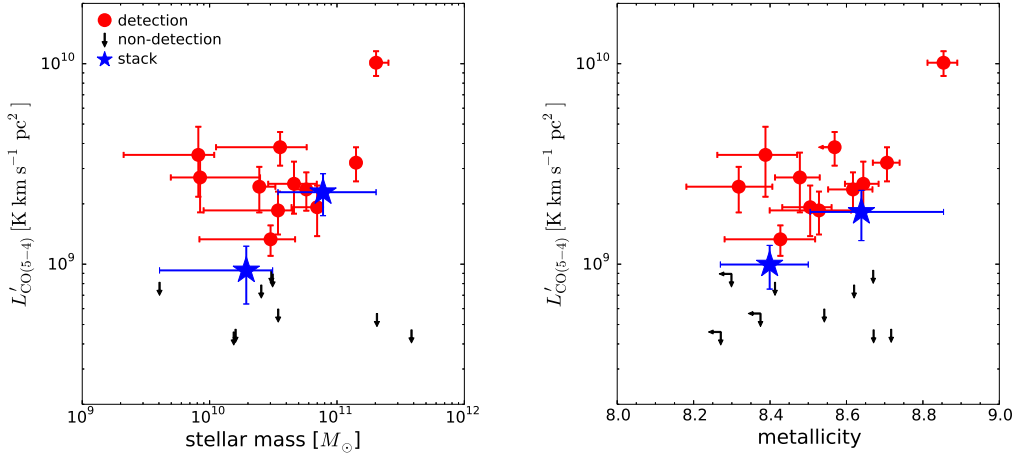}
		\caption{CO(5-4) luminosity plotted against stellar mass
                  (Left) and metallicity (Right).  Filled
                  (red) circles refer to the CO detected galaxies and
                  arrows show the upper limits.  Filled (blue) stars
                  refer to the results of stacking analysis for the
                  subsamples with larger/smaller stellar mass in left
                  panel (see the section \ref{subsubsec: stack mass
                    CO}) and the subsamples with higher/lower
                  metallicity in right panel (see the section
                  \ref{subsubsec: stack metal CO}). }
\label{fig: Lco54}
\end{center}
\end{figure*}

The CO($J=5-4$) line luminosity ($L_\mathrm{CO(5-4)}^{'}$) is given as
\begin{equation}
L_\mathrm{CO(5-4)}^{'} = 3.25 \times 10^{7} S_\mathrm{CO(5-4)} \Delta v \nu_\mathrm{rest(5-4)}^{-2} D_{L}^2 (1+z)^{-1}, 
\end{equation}
where $L_\mathrm{CO(5-4)}^{'}$ is measured in
$\mathrm{K~km~s^{-1}~pc^{2}}$, $S_\mathrm{CO(5-4)} \Delta v$ is the
observed CO(5-4) integrated flux density in $\mathrm{Jy~km~s^{-1}}$,
$\nu_\mathrm{rest(5-4)}$ is the rest frequency of the CO(5-4) emission
line in GHz, and $D_L$ is the luminosity distance in $\mathrm{Mpc}$.
For the non-detected galaxies, we make channel maps with a velocity
resolution of 200~$\mathrm{km~s^{-1}}$, and measure the noise levels
($\sigma_{200}$), because the FWHM of the detected CO emission lines
range from $45~\mathrm{km~s^{-1}}$ to $490~\mathrm{km~s^{-1}}$ and the
average FWHM is about $200~\mathrm{km~s^{-1}}$.  We take a
2$\sigma_{200}$ upper limit for the CO(5-4) flux density and a
velocity width of 200~$\mathrm{km~s^{-1}}$.  CO(5-4) luminosities and
upper limits are shown in Table \ref{table: CO} and plotted against
stellar mass and metallicity in Figure \ref{fig: Lco54}.  The CO
luminosities of the detected galaxies (filled red circles) appear to
increase with increasing stellar mass and metallicity.

\begin{figure*}
\begin{center}
	\epsscale{2.0}
		\plotone{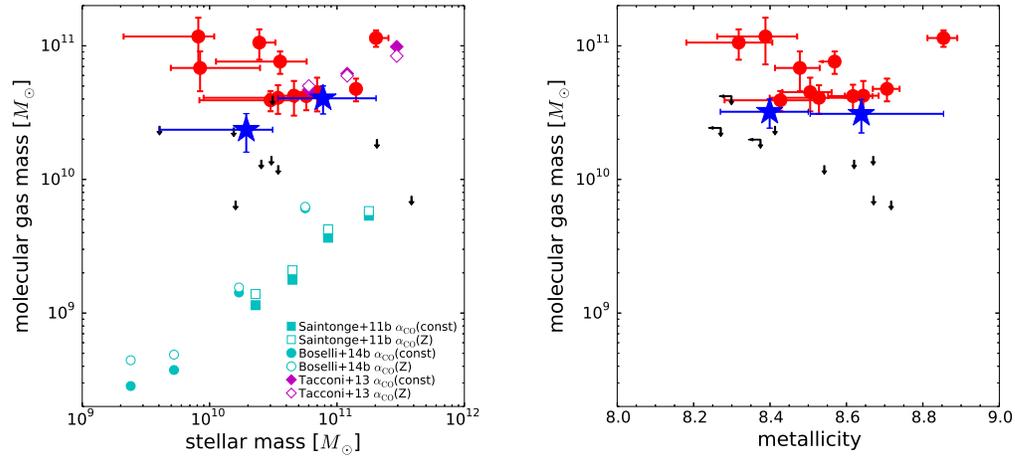}
		\caption{Molecular gas mass plotted against stellar mass
                  (Left) and metallicity (Right).  We adopted
                  a CO(5-4)/CO(1-0) luminosity ratio of 0.23 and the
                  metallicity-dependent CO-to-H$_2$ conversion factors
                  shown in the equation \ref{eq: conversion factor}.
                  Filled (red) circles, arrows, and filled (blue) stars are 
                  the same as those in Figure \ref{fig: Lco54}. 
                  Filled (cyan) squares and circles refer to the average values 
                  in local galaxies with a constant CO-to-H$_2$ conversion factor 
                  given by \citet{Sain11a} and \citet{Bose14b}, respectively. 
                  Filled (magenta) diamonds refer to the average values 
                  at $z\sim1.5$ with the Galactic conversion factor \citep{Tacc13}. 
                  Open (cyan and magenta) symbols show the values obtained 
                  with a metallicity-dependent CO-to-H$_2$ conversion factor. 
                  See text for more details. 
                  }
\label{fig: Mgas}
\end{center}
\end{figure*}

\begin{figure*}
\begin{center}
	\epsscale{2.0}
		\plotone{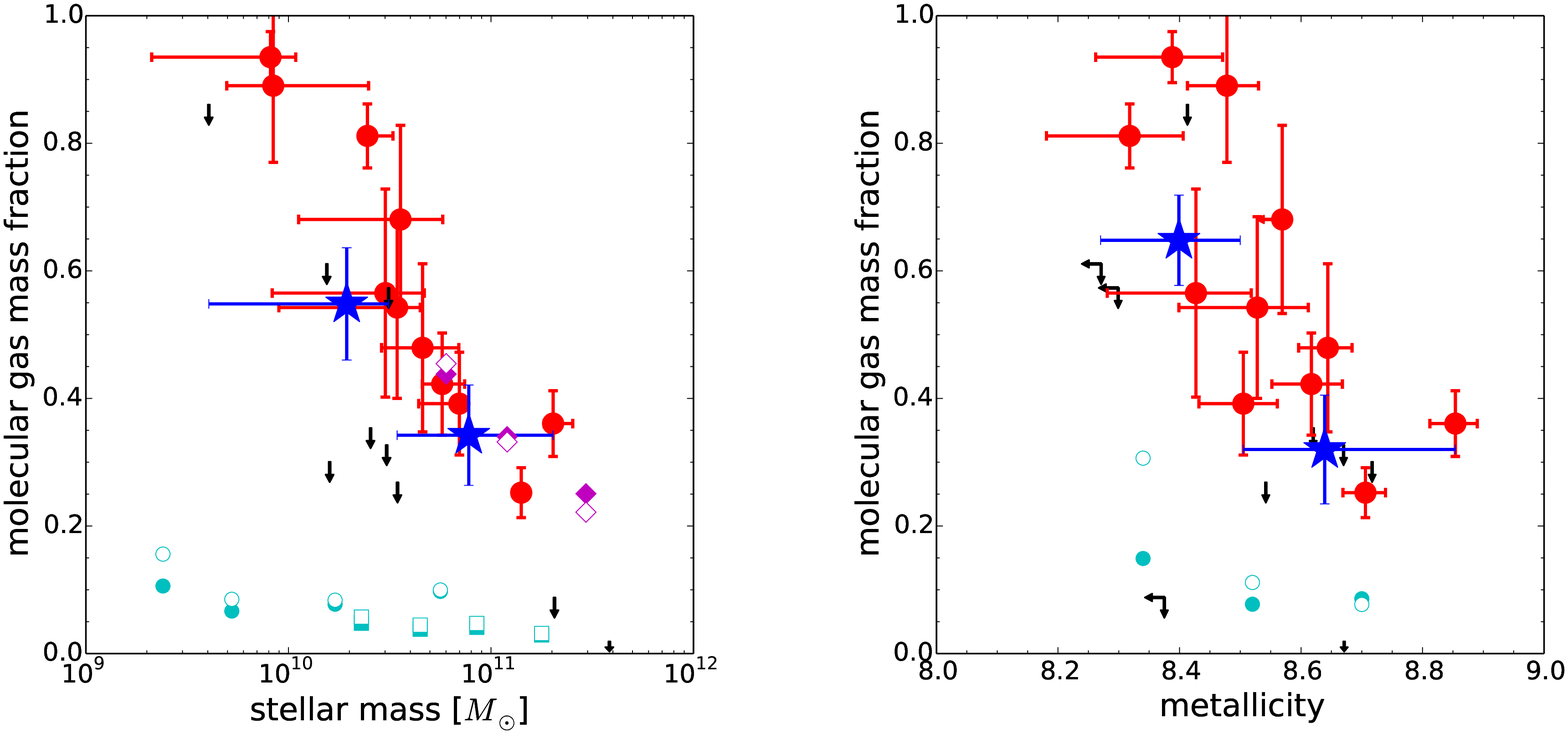}
		\caption{Molecular gas mass fraction plotted against
                  stellar mass (Left) and metallicity (Right).
                  Symbols are the same as those in Figure \ref{fig: Mgas}. }
\label{fig: fgas}
\end{center}
\end{figure*}

The molecular gas mass is derived from
\begin{equation}
M_\mathrm{mol} = \alpha_\mathrm{CO} L_\mathrm{CO(1-0)}^{'}. 
\end{equation}
To derive the molecular gas mass, CO(5-4)/CO(1-0) luminosity ratio is
needed.  According to a study of the luminosity ratios in three sBzK
galaxies at $z\sim1.5$ \citep{Dadd15}, the average CO(5-4)/CO(1-0)
luminosity ratio is 0.23, corresponding to $S_\mathrm{CO(5-4)} \Delta
v/S_\mathrm{CO(1-0)} \Delta v \sim 6$, with an uncertainty of a factor
of 2.  We adopt this value for the conversion of CO(5-4) luminosity to
CO(1-0) luminosity.  In local galaxies the value of
$\alpha_\mathrm{CO}$ correlates with gas metallicity: the value of
$\alpha_\mathrm{CO}$ is larger in galaxies with lower metallicity
\citep[e.g.,][]{Arim96,Lero11}.  A similar relation is found in
star-forming galaxies at $z=1-2$ \citep{Genz12}.  We adopt the
equation (7) by \citet{Genz12}:
\begin{equation} \label{eq: conversion factor}
\log(\alpha_\mathrm{CO}) = -1.3 \times (12 + \log(\mathrm{O/H}))_\mathrm{Denicol\acute{o}\ 02} + 12, 
\end{equation}
where $12 + \log(\mathrm{O/H})_\mathrm{Denicol\acute{o}\ 02}$ is
metallicity calibrated by \citet{Deni02}.  Since we use the
metallicity calibration of \citet{PP04}, we convert the metallicity
using an empirical relation between the two metallicity calibrations
of \citet{Kewl08}.  Derived molecular gas masses are listed in Table
\ref{table: CO} and plotted against stellar mass and metallicity in
Figure \ref{fig: Mgas}.  The uncertainty given to the molecular gas
mass is based on SN of the integrated intensity map and does not
include the uncertainty of the luminosity ratio and
$\alpha_\mathrm{CO}$ (uncertainty of a factor of $\sim2$).  For the
detected galaxies (filled red circles), the molecular gas mass does
not seem to depend on stellar mass and metallicity.

The derived molecular gas mass fractions against stellar masses are
also listed in Table \ref{table: CO} and plotted against stellar mass
and metallicity in Figure \ref{fig: fgas}.  For the detected galaxies
(filled red circles), the molecular gas mass fraction decreases with
increasing stellar mass.  This trend is the same in the previous
studies \citep[e.g.,][]{Tacc13}.  We show that the trend holds in
galaxies with lower stellar mass than have been observed in previous
studies.  Furthermore, we found that the molecular gas mass fraction
decreases with increasing metallicity.

\subsection{Stacking analysis}
Since the CO emission lines from about half of our sample galaxies are
not detected, we carried out a stacking analysis to examine the
relations against stellar mass and metallicity.  For the stacking
analysis, we use images without applying cleaning (i.e., dirty maps).
The images are stacked on a pixel by pixel basis using weighted
average.  The weights are calculated as $1/\sigma^{2}$, where $\sigma$
is the R.M.S. noise level in each map.  \citet{Lind15} constructed an
algorithm for $uv$-stacking, and they found that image-stacking
produced similar results to $uv$-stacking.  To study the properties of
the ISM in main sequence galaxies, we did not include the two most
massive galaxies, SXDS1\_35572 and SXDS1\_79307, in our stacking
analysis.

For the stacking analysis of CO emission lines, we used channel maps
with $50~\mathrm{km~s^{-1}}$ bins over a range of
$\pm1000~\mathrm{km~s^{-1}}$ from the zero velocities.  Then, we
stacked the maps at the same velocity.  The detection criteria for the
stacked maps are the same as those for the individual galaxies.  Error
bars of integrated intensity are derived from a random resampling of
stacked galaxies.  The molecular gas mass is derived using the same CO
luminosity ratio and metallicity-dependent CO-to-H$_2$ conversion
factor as those used for individual galaxies.  The stellar mass and
redshift of the stacks are taken to be average values of
the stacked galaxies.  The metallicity is also derived from stacked FMOS
spectra \citep[detailes are described in][]{Yabe12, Yabe14}.

\begin{figure*}
\begin{center}
	\epsscale{2.0}
		\plotone{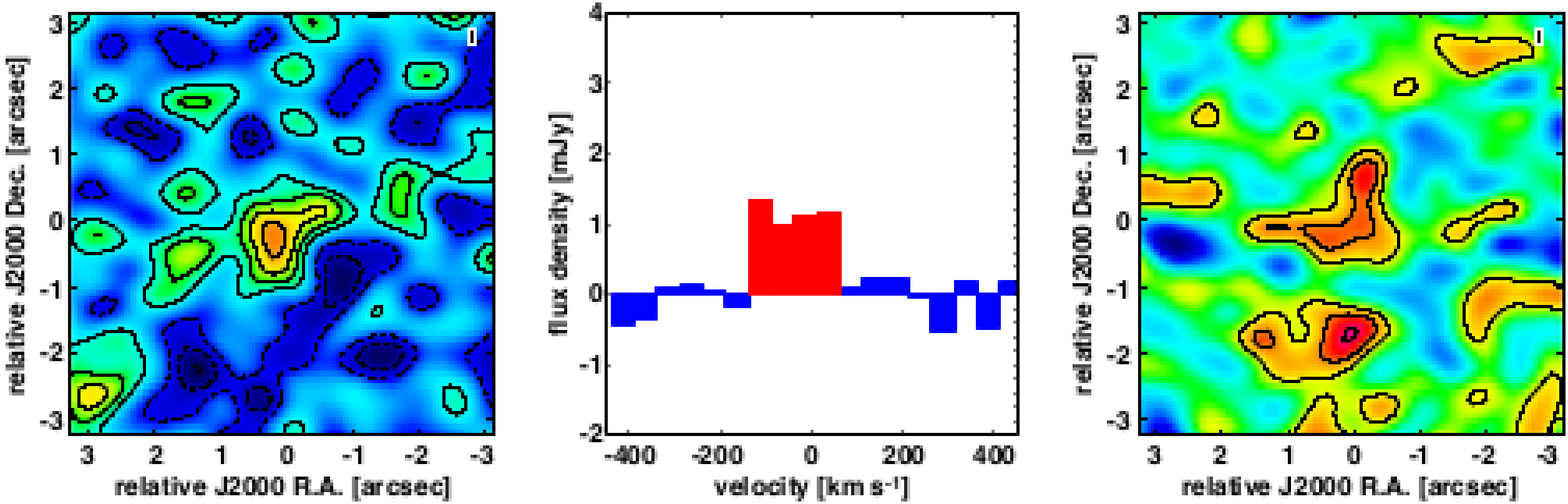}
		\plotone{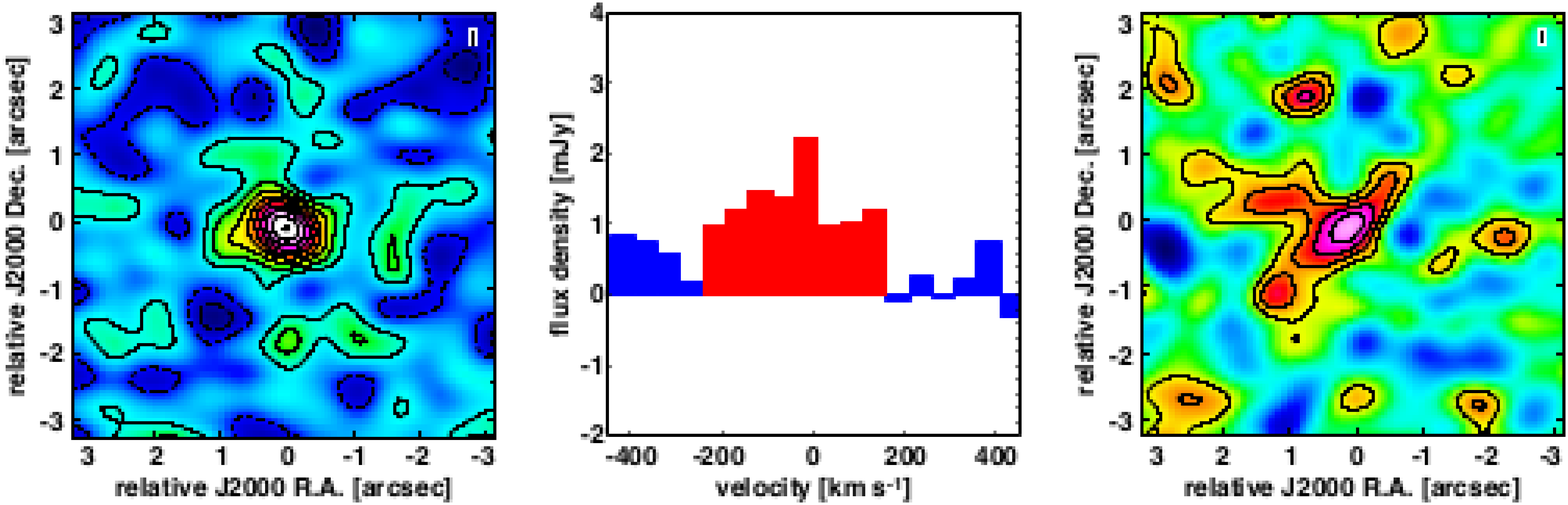}
		\plotone{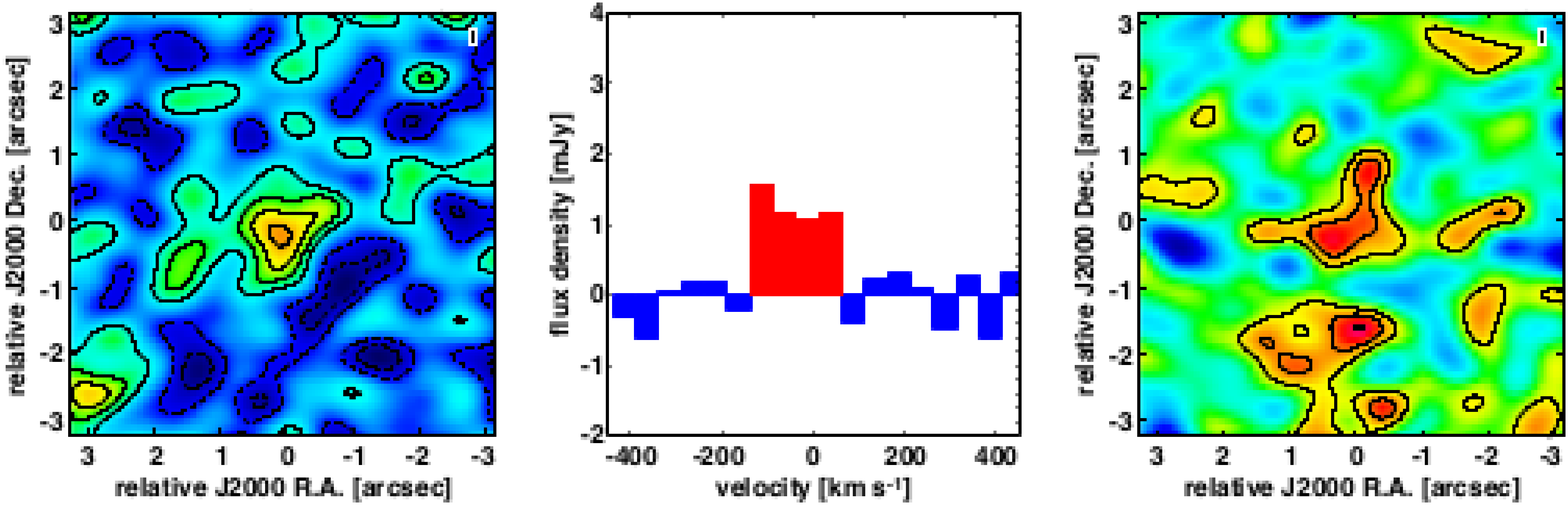}
		\plotone{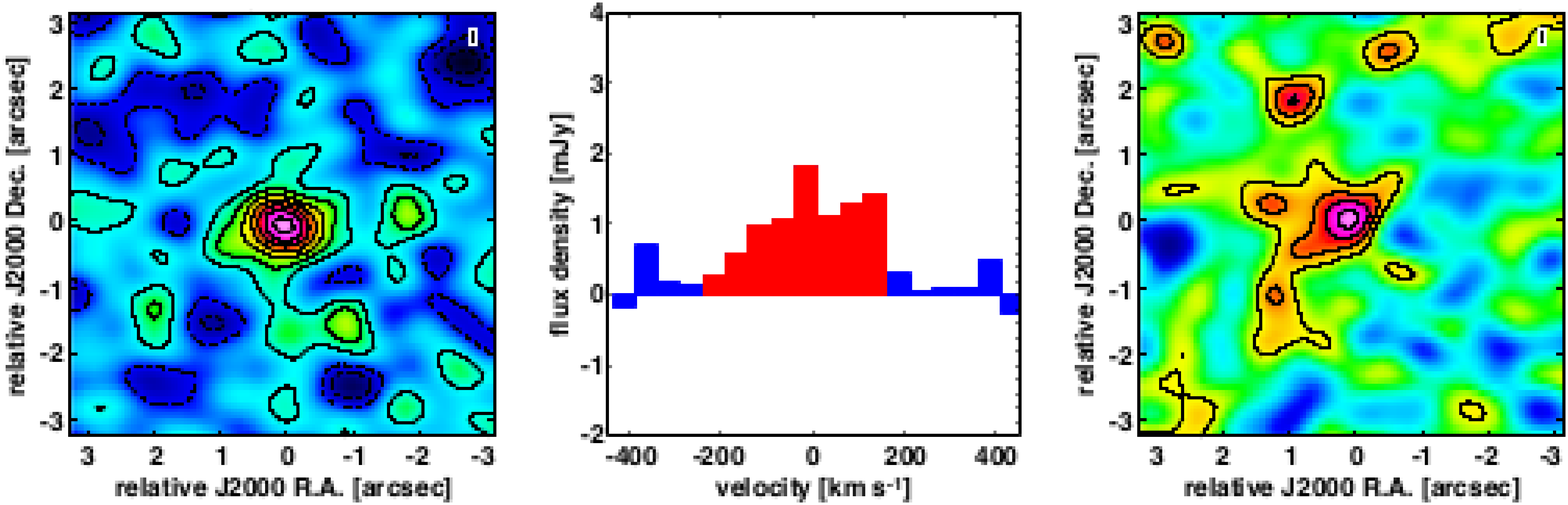}
		\caption{(Top left) Stacked integrated CO(5-4)
                  intensity map of the smaller stellar mass subsample
                  (section \ref{subsubsec: stack mass CO}).  Contours
                  represent $-2\sigma$, $-1\sigma$ (dashed lines),
                  $1\sigma$, $2\sigma$, $3\sigma$, \ldots (solid lines).
                  (Top center) Stacked CO(5-4) spectrum of the smaller
                  stellar mass subsample.  CO emission is shown with
                  red color which also shows the velocity range to
                  make the intensity map.  The zero velocity is
                  derived by referring the spectroscopic redshift of
                  H$\alpha$ emission.  The spectrum is made in the
                  region where SN is $>1$ around the stacked
                  source.  (Top right) Stacked continuum map of the
                  smaller stellar mass subsample (section
                  \ref{subsubsec: stack mass dust}).  Contours
                  represent $1\sigma$, $2\sigma$, $3\sigma$,
                  \ldots (solid lines).  (Upper middle) Same as the top
                  row, but for the stacking analysis of the
                  larger stellar mass subsample (section
                  \ref{subsubsec: stack mass CO} and \ref{subsubsec:
                    stack mass dust}).  (Lower middle) Same as the top
                  row, but for the stacking analysis of the
                  lower metallicity subsample (section \ref{subsubsec:
                    stack metal CO} and \ref{subsubsec: stack metal
                    dust}).  (Bottom) Same as the top row, but for the
                  stacking analysis of the higher
                  metallicity subsample (section \ref{subsubsec: stack
                    metal CO} and \ref{subsubsec: stack metal dust}).
                }
\label{fig: stack image}
\end{center}
\end{figure*}

\begin{figure*}[!H]
\begin{center}
	\epsscale{2.0}
		\plotone{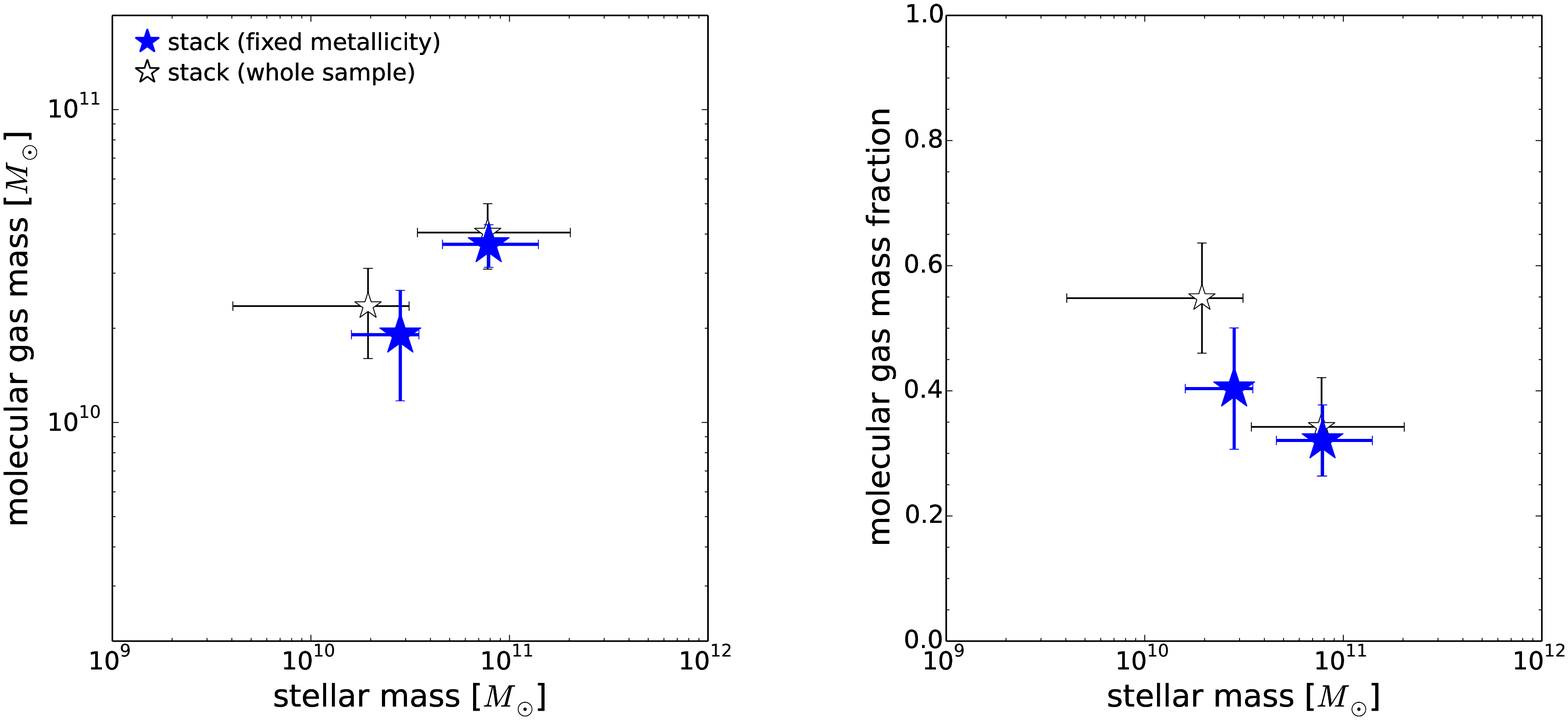}
		\plotone{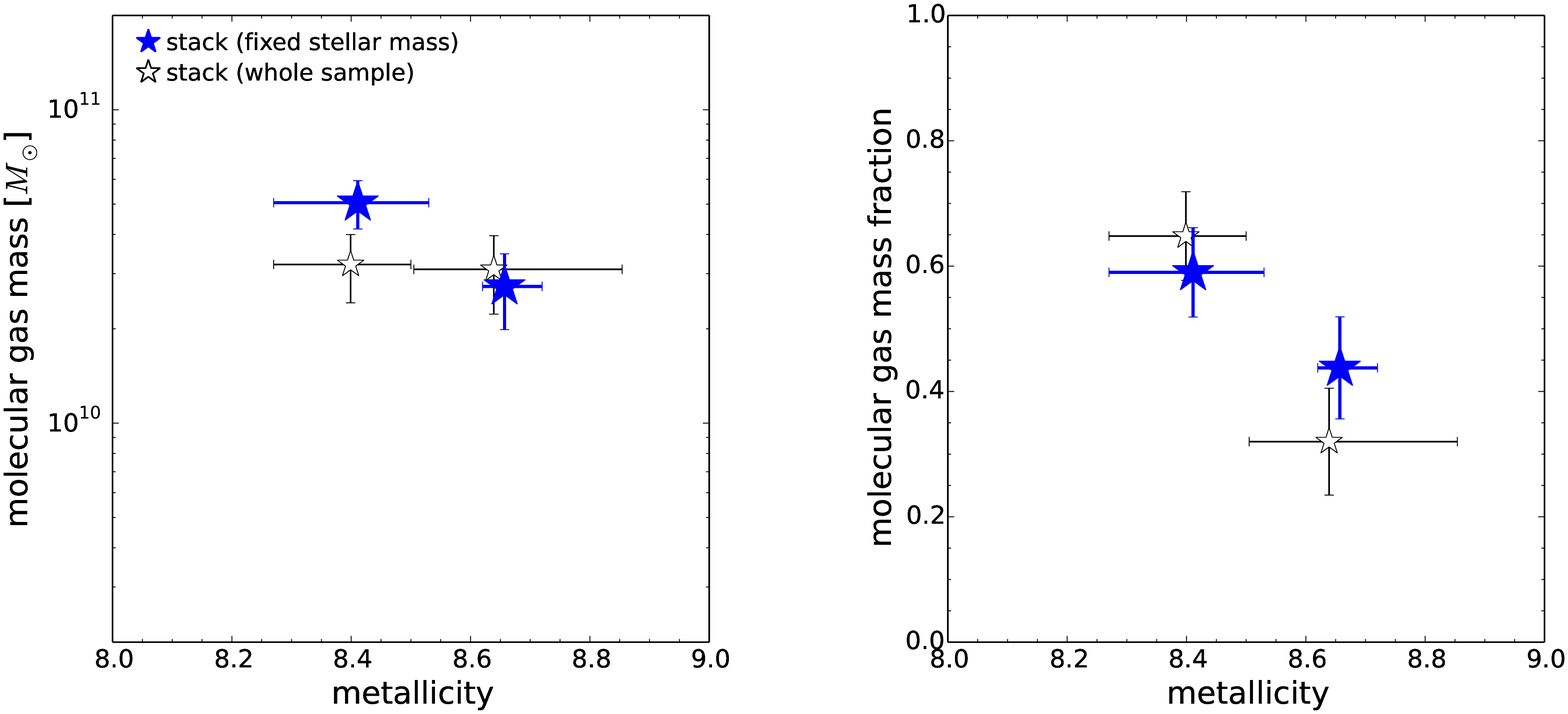}
		\caption{Molecular gas mass (upper left) and molecular
                  gas mass fraction (upper right) against stellar
                  mass.  Large filled blue stars refer to the stacking
                  analysis for the subsamples with larger/smaller
                  stellar mass but with almost the same stacked
                  metallicity.  Small open stars refer to the stacking
                  analysis for the subsamples with larger/smaller
                  stellar mass using the whole sample.  Molecular gas
                  mass (bottom left) and molecular gas mass fraction
                  (bottom right) against metallicity.  Large filled
                  blue stars refer to the stacking analysis for the
                  subsamples with higher/lower metallicity but with
                  the same average stellar mass.  Small open stars
                  refer to the stacking analysis for the subsamples
                  with higher/lower metallicity using the whole
                  sample. }
\label{fig: stack discussion}
\end{center}
\end{figure*}

\subsubsection{Stacking analysis of subsamples with larger/smaller stellar mass} \label{subsubsec: stack mass CO}
In order to examine the dependences of molecular gas mass and gas mass
fraction on stellar mass, we carried out a stacking analysis for
subsamples with smaller stellar mass
($(0.4-3.1)\times10^{10}~M_\odot$, 10 galaxies) and larger stellar
mass ($(3.4-20)\times10^{10}~M_\odot$, 8 galaxies).  The noise levels
in the channel maps for these subsamples are similar
($\sim0.25~\mathrm{mJy~beam^{-1}}$ at a velocity resolution of
$50~\mathrm{km~s^{-1}}$).  The resulting stacked images for subsamples
with smaller and larger stellar mass are shown in left
panel of the top and upper middle row of Figure \ref{fig: stack
  image}, respectively.  The stacked profiles are shown in the middle
panel of Figure \ref{fig: stack image}.  The integrated intensity maps
are made with the velocity range shown with red in the profiles.  The
CO emissions are significantly detected for both subsamples.
CO($5-4$) line luminosities are
$(9.3\pm3.0)\times10^{8}~\mathrm{K~km~s^{-1}~pc^{2}}$ and
$(2.3\pm0.5)\times10^{9}~\mathrm{K~km~s^{-1}~pc^{2}}$ for the smaller
and larger stellar mass subsamples, respectively, and are plotted
against stellar mass in the left panel of Figure \ref{fig: Lco54} (filled
blue stars).  
We carried out the Welch's $t$ test to evaluate the significance of difference 
between the stacked values of the subsamples. 
Here, null hypothesis is that mean values of both subsamples are equal. 
Significance level for null hypothesis is 0.003\%.
Thus we regard
the CO luminosity significantly increases with
increasing stellar mass.  
The resulting molecular gas masses of the
smaller and larger stellar mass subsamples are
$(2.4\pm0.7)\times10^{10}~M_\odot$ and
$(4.0\pm1.0)\times10^{10}~M_\odot$, respectively, and are plotted in the
left panel of Figure \ref{fig: Mgas} (filled blue stars).  

We compare the results with those in local galaxies 
(CO Legacy Database for Galex Arecibo SDSS Survey 
\citep[COLD GASS:][]{Sain11a} 
and {\it Herschel} Reference Survey \citep[HRS:][]{Bose14a}). 
Because the studies in local galaxies used Chabrier IMF, 
we converted the stellar mass and SFR to those 
with Salpeter IMF by multiplying 1.7 \citep{Spea14}. 
In addition, the CO-to-H$_2$ conversion factor used in the studies 
does not include the helium mass and is not corrected 
for metallicity dependence. 
Thus, in the left panel of Figure~\ref{fig: Mgas}, 
we also plot the local molecular gas mass considering 30\% contribution 
of helium (filled cyan symbols) and the metallicity-dependent 
CO-to-H$_2$ conversion factor \citep[open cyan symbols:][]{Lero11, Bola13}. 
We estimated the metallicity of local sample from the 
mass$-$metallicity relation in local universe whose 
metallicity is derived based on N2 method \citep{Erb06}; 
\citet{Erb06} derived the local relation with N2 method 
by using Sloan Digital Sky Survey (SDSS) galaxies 
to compare the relation at $z\sim2$.  
The study of molecular gas at $z\sim1.5$ by \citet{Tacc13} also used 
Chabrier IMF and the Galactic CO-to-H$_2$ conversion factor 
($\sim4.36~M_\odot~(\mathrm{K~km~s^{-1}~pc^2})^{-1}$; including helium mass). 
We converted the stellar mass and SFR to those with Salpeter IMF 
and plotted the molecular gas mass in the left panel 
of Figure~\ref{fig: Mgas} (filled magenta diamonds) 
and plotted that derived with the metallicity-dependent 
CO-to-H$_2$ conversion factor given by the equation~(\ref{eq: conversion factor}) (open magenta diamonds).
The metallicity was estimated from the mass$-$metallicity relation 
at $z\sim1.4$ \citep{Yabe14}. 
The molecular gas masses of our sample galaxies are significantly 
larger than those in local star-forming galaxies with similar stellar mass
\citep[$M_\mathrm{mol}\sim3\times10^{9}~M_\odot$;][]{Sain11a}.  
The molecular gas mass seems to increase with increasing stellar mass
(significance level for null hypothesis is 0.25\%).

The gas mass fractions are $0.55\pm0.09$ and $0.34\pm0.08$ for the
smaller and larger stellar mass subsamples, respectively, and are
plotted against stellar mass in the left panel of Figure \ref{fig: fgas}
(filled blue stars).  These gas mass fractions are also significantly
larger than those in local star-forming galaxies
\citep[$f_\mathrm{mol}\sim0.08$;][]{Sain11a}.  The gas fraction
significantly decreases with increasing stellar mass
(significance level for null hypothesis is 0.01\%).  
This trend is the same as that in local galaxies \citep[e.g.,][]{Sain11a, Bose14b} 
and in previous studies at similar redshift \citep[e.g.,][]{Tacc13}, but our
sample extends to the lower stellar mass.

\subsubsection{Stacking analysis of subsamples with higher/lower metallicity} 
\label{subsubsec: stack metal CO}
In order to examine the dependences of molecular gas mass and gas mass
fraction on metallicity, we made stacking analysis for subsamples with
lower metallicity ($12+\log(\mathrm{O/H})<8.5$, 7 galaxies) and higher
metallicity ($12+\log(\mathrm{O/H})>8.5$, 10 galaxies).  We exclude
SXDS1\_42087 because the metallicity is an upper limit
($12+\log(\mathrm{O/H})<8.57$).  The noise levels in the channel maps of
these subsamples are similar ($\sim0.25~\mathrm{mJy~beam^{-1}}$ at a
velocity resolution of $50~\mathrm{km~s^{-1}}$).  The resulting stacked
images for subsamples with lower metallicity and higher metallicity
are shown in left panel of the lower middle and bottom rows of Figure
\ref{fig: stack image}, respectively.  The stacked profiles are shown
in the middle panel of Figure \ref{fig: stack image}.  The integrated
intensity maps are made with the velocity range shown in red in the
profiles.  The CO emissions are significantly detected for both
subsamples.  CO($5-4$) line luminosities are
$(1.0\pm0.3)\times10^{9}~\mathrm{K~km~s^{-1}~pc^{2}}$ and
$(1.8\pm0.5)\times10^{9}~\mathrm{K~km~s^{-1}~pc^{2}}$ for the lower
and higher metallicity subsamples, respectively, and are plotted
against metallicity in the right panel of Figure \ref{fig: Lco54} (filled
blue stars).  The CO luminosity seems to increase with increasing
metallicity (significance level for null hypothesis is 0.1\%).
The resulting molecular gas masses of the lower and
higher metallicity subsamples are $(3.2\pm0.8)\times10^{10}~M_\odot$
and $(3.1\pm0.9)\times10^{10}~M_\odot$, respectively, and are plotted
in the right panel of Figure \ref{fig: Mgas} (filled blue stars).  The
molecular gas mass does not depend on metallicity
(significance level for null hypothesis is 82\%).  
In this Figure, the result in local
star-forming galaxies is not shown. Because \citet{Bose14b} 
showed the molecular gas mass-to-stellar mass ratio 
but did not show the stellar mass for the metallicity-based analysis,
we were not able to calculate the molecular gas mass. 

The gas mass fractions are $0.65\pm0.07$ and $0.32\pm0.09$ 
for the lower and higher metallicity subsamples, respectively, 
and are plotted against metallicity in the right panel of Figure \ref{fig: fgas} 
(filled blue stars).  The molecular gas mass fraction significantly decreases 
with increasing metallicity (significance level for null hypothesis is less than 0.001\%).

\subsection{Relations for subsamples with fixed metallicity and fixed stellar mass}
The relations between gas mass or its fraction and stellar mass or
metallicity are examined.  Due to the mass-metallicity relation,
however, the dependences on stellar mass and metallicity are not
clearly separated.  Thus we next investigate the dependence only on
stellar mass and on metallicity by using a stacking analysis.

First, to avoid the metallicity effect, we made the stacking analysis
for subsamples with smaller ($(1-4)\times10^{10}~M_\odot$) and larger
($(4-15)\times10^{10}~M_\odot$) stellar mass but with almost the same
metallicity ($8.50-8.75$).  The subsamples with smaller and larger
stellar mass include 5 and 4 galaxies, respectively.  We exclude
SXDS1\_42087 because the metallicity is an upper limit of
$12+\log(\mathrm{O/H})<8.57$.  The average stellar mass and stacked
metallicity of the smaller stellar mass subsample are
$2.8\times10^{10}~M_\odot$ and 8.61 and those of larger stellar mass
subsample are $7.9\times10^{10}~M_\odot$ and 8.63.  According to the
mass-metallicity relation at $z\sim1.4$ \citep[e.g.,][]{Yabe14}, the
difference of stellar mass in these subsamples
($2.8\times10^{10}~M_\odot$ and $7.9\times10^{10}~M_\odot$) produces
the difference of metallicity of 0.07.  Thus, the difference of
metallicity in these subsamples (8.61 and 8.63) does not trace the
mass-metallicity relation and it is reasonable to consider that only
the stellar mass effect can be seen.  The resulting molecular gas masses
of the smaller and larger stellar mass subsamples are
$(1.9\pm0.7)\times10^{10}~M_\odot$ and
$(3.7\pm0.6)\times10^{10}~M_\odot$, respectively.  We plot the result
in the upper left panel of Figure \ref{fig: stack discussion} (filled
blue stars).  The trend for the gas mass (significance level for null 
hypothesis is 0.4\%) is the same as that with whole sample galaxies (open stars).  
The gas mass fractions are
$0.40\pm0.10$ and $0.32\pm0.06$ for the smaller and larger stellar
mass subsamples, respectively.  We plot the result in the upper right
panel of Figure \ref{fig: stack discussion} (filled blue stars).
Although the trend for the gas mass fraction seems to be the same as
that for whole sample (open stars), it is not so significant 
(significance level for null hypothesis is 18\%).

Next, to avoid the stellar mass effect, we made the stacking analysis
for subsamples with lower ($<8.55$) and higher ($8.6-8.8$) metallicity
but with comparable stellar mass ($10^{10-11}~M_\odot$).  The
subsamples with lower and higher metallicity both include 5 galaxies.
The average stellar mass and stacked metallicity of the lower
metallicity subsample are $3.5\times10^{10}~M_\odot$ and 8.41, and
those of higher metallicity subsample are $3.5\times10^{10}~M_\odot$
and 8.66.  Since there is no difference of stellar mass in these
subsamples, it is reasonable to conclude that only the metallicity effect
can be seen.  The resulting molecular gas masses of the lower and
higher metallicity subsamples are $(5.3\pm0.9)\times10^{10}~M_\odot$
and $(2.8\pm0.8)\times10^{10}~M_\odot$, respectively.  We plot the
results in the bottom left panel of Figure \ref{fig: stack discussion}
(filled blue stars).  The molecular gas mass decreases with increasing
metallicity (significance level for null hypothesis is 0.2\%); 
this trend is different from that for the whole sample of galaxies (open stars). 
The gas mass fractions are $0.60\pm0.07$ and
$0.45\pm0.08$ for the lower and higher metallicity subsamples,
respectively.  We also plot the result in the bottom right panel of
Figure \ref{fig: stack discussion} (filled blue stars).  The gas mass
fraction seems to decrease with metallicity 
(significance level for null hypothesis is 1.3\%), 
which is the same trend as seen for the whole sample galaxies (open stars).

\begin{figure*}
\begin{center}
	\epsscale{2.0}
		\plotone{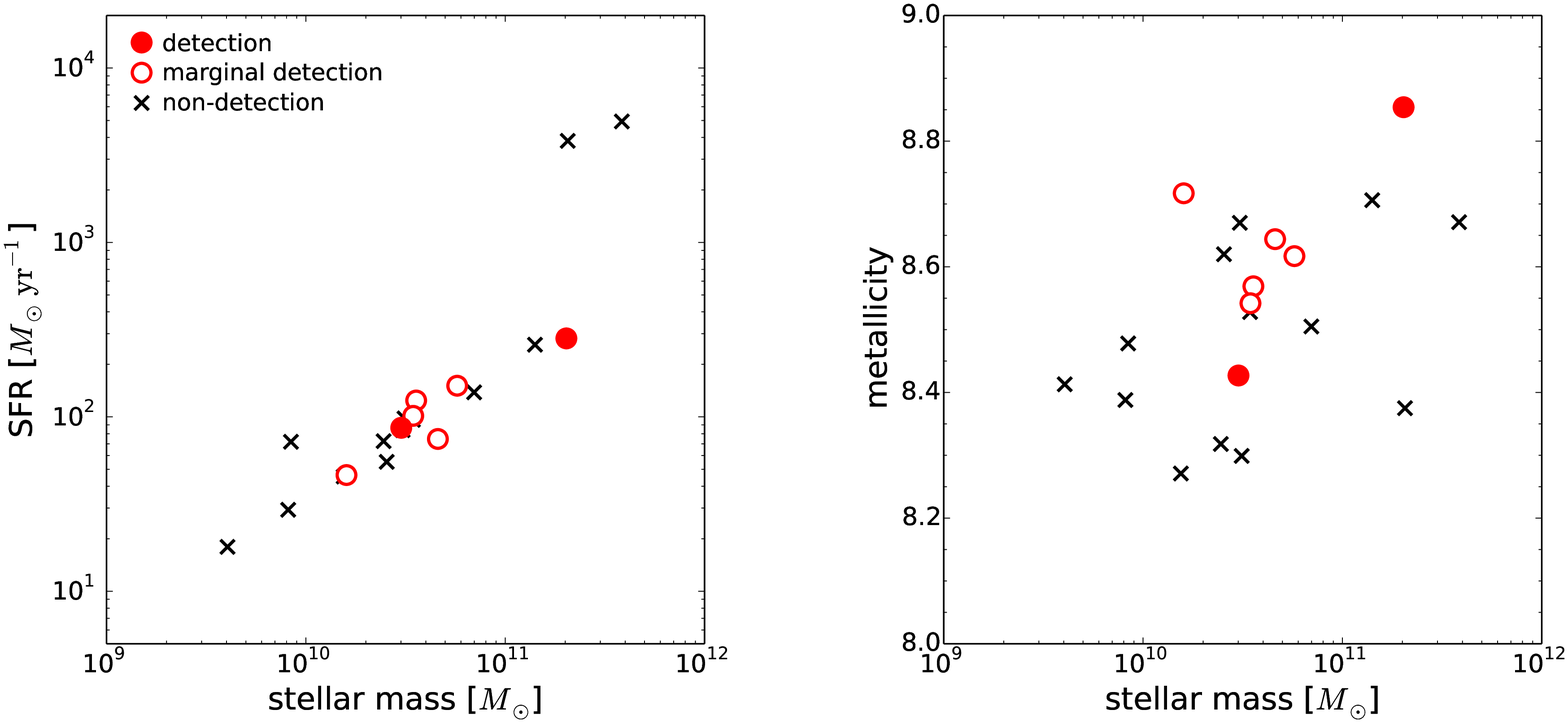}
		\caption{Observed galaxies in the stellar mass$-$SFR
                  diagram (Left) and in the stellar mass$-$metallicity
                  diagram (Right).  Filled and open (red) circles
                  refer to the galaxies for which the continuum
                  emissions are significantly and marginally detected,
                  respectively.  The crosses show non-detections. }
\label{fig: continuum detected sources}
\end{center}
\end{figure*}

\begin{figure*}
\begin{center}
	\epsscale{2.0}
		\plotone{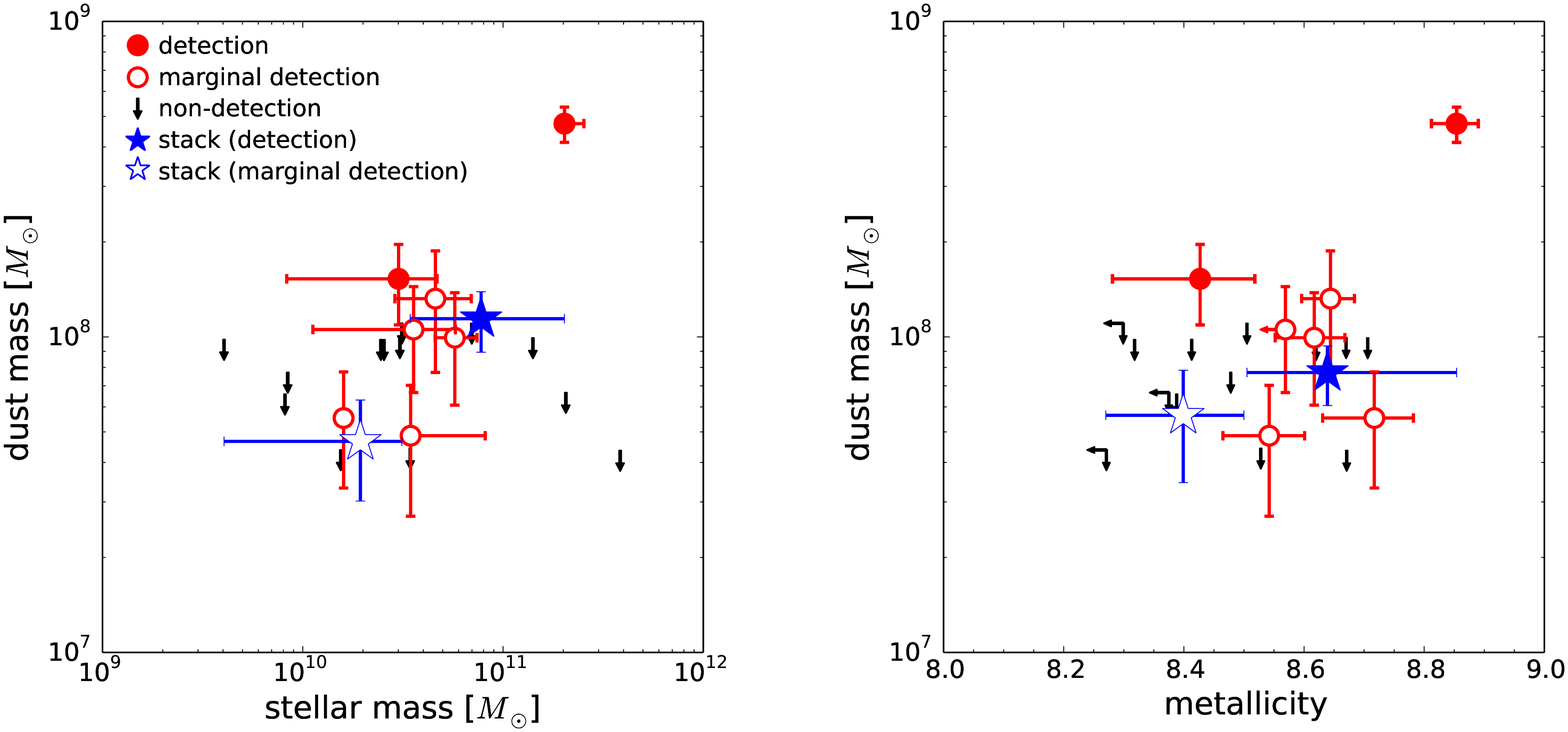}
		\caption{Dust mass against stellar mass (Left) and
                  against metallicity (Right).  Dust masses are derived
                  using a modified black body model by assuming a dust
                  temperature of 30~K and a dust emissivity index of
                  1.5.  The filled and open (red) circles refer to the
                  galaxies for which the continuum emissions are
                  significantly and marginally detected, respectively.
                  Arrows show the upper limits.  Filled/open (blue)
                  stars refer to the results of the stacking analyses for
                  the subsamples with larger/smaller stellar mass in
                  left panel (see the section \ref{subsubsec: stack
                    mass dust}) and the subsamples with higher/lower
                  metallicity in right panel (see the section
                  \ref{subsubsec: stack metal dust}). }
\label{fig: Mdust}
\end{center}
\end{figure*}

\section{Results for dust thermal emissions} \label{sec: result dust}
\subsection{Individual galaxy}
We detected continuum emission from two galaxies with $\mathrm{SN} > 3$.  For
five objects, the emissions are marginally detected with $2 <
\mathrm{SN} < 3$.  The positions of these galaxies are in good
agreement with the centroid of each galaxy in the K-band image.  The
detected and marginally detected galaxies are shown with filled
circles and open circles, respectively, in the stellar mass$-$SFR and
stellar mass$-$metallicity diagrams (Figure \ref{fig: continuum
  detected sources}).  Continuum emission seems to tend to be detected
for more massive galaxies and galaxies with higher metallicity.

The continuum emission is considered to originate from dust
thermal emission.  The dust mass ($M_\mathrm{dust}$) is derived as
\begin{equation} \label{eq: Mdust}
M_\mathrm{dust} = \frac{S_\mathrm{cont} D_\mathrm{L}}{(1+z) \kappa_\mathrm{d}(\nu_\mathrm{rest}) B(\nu_\mathrm{rest}, T_\mathrm{dust})}, 
\end{equation}
where $S_\mathrm{cont}$ is the observed flux density of dust thermal
continuum emission, $\kappa_\mathrm{d}(\nu_\mathrm{rest})$ is the dust
mass absorption coefficient in the rest-frame frequency 
($\sim570~\mathrm{GHz}$; rest-frame wavelength is $\sim0.5~\mathrm{mm}$),
$T_\mathrm{dust}$ is the dust temperature, and $B(\nu_\mathrm{rest},
T_\mathrm{dust})$ is the Planck function.  $\kappa_\mathrm{d}$ varies
with frequency as $\kappa_\mathrm{d} \propto \nu^{\beta}$, where
$\beta$ is the dust emissivity index.  We adopt
$\kappa_\mathrm{d}(125~\mu\mathrm{m}) = 1.875~\mathrm{m^2~kg^{-1}}$
\citep{Hild83}, and $\beta = 1.5$.  \citet{Magn14} derived dust
temperatures of star-forming galaxies in the stellar mass$-$SFR
diagram at $z=0-2.3$.  According to their results, the dust
temperatures of main sequence galaxies at $z=1.2-1.7$ are
$25-35~\mathrm{K}$.  We adopt a temperature of $30~\mathrm{K}$.  
We take a $2\sigma$ upper limit as $S_\mathrm{cont}$, where 
$\sigma$ is the R.M.S. noise level in continuum map. 
The derived dust masses are given in Table \ref{table: CO} and plotted
against stellar mass and metallicity in Figure \ref{fig: Mdust}.  The
uncertainty in the dust mass is calculated from the SN of the
continuum map.  The dust mass can change by a factor of 1.2 when we
adopt dust temperatures of $25\ \mathrm{or}\ 35$~K with a dust
emissivity index of 1.5, and change by a factor of 2 when we adopt a
dust emissivity indices of $1.0\ \mathrm{or}\ 2.0$ with a dust
temperature of 30~K.  No clear dependence on stellar mass or
metallicity is seen.

\subsection{Stacking analysis} \label{sec: stack}
Since the dust thermal continuum emissions from most of the observed
galaxies are not detected, we carried out a stacking analysis to
examine the dependence on stellar mass and metallicity.  For the
stacking analysis of dust thermal emission we used uncleaned images.
The images are stacked on a pixel by pixel basis using $1/\sigma^{2}$
weighted average.  We do not include SXDS1\_35572 and SXDS1\_79307 as
discussed previously.  The detection criteria are the same as those
for the individual galaxies.  The dust mass is derived with the same
dust temperature and dust emissivity index as for the individual
galaxies.  Error bars are derived from a random resampling of the
stacked galaxies.

\subsubsection{Stacking analysis of subsamples with larger/smaller stellar mass} \label{subsubsec: stack mass dust}
To study the stellar mass dependence of dust mass, we carried out a
stacking analysis with the subsamples used in section \ref{subsubsec:
  stack mass CO}.  The stacked continuum maps for the subsamples with
smaller and larger stellar mass are shown in the right panel of the
top and upper middle rows of Figure \ref{fig: stack image},
respectively.  The dust emission is marginally detected for the
subsamples with smaller stellar mass, and significantly detected for
the subsamples with larger stellar mass.  The derived dust masses are
$(4.7\pm1.6)\times10^{7}~M_\odot$ and
$(1.1\pm0.3)\times10^{8}~M_\odot$ for the subsamples with smaller and
larger stellar mass, respectively, and are plotted with blue stars in
the left panel of Figure \ref{fig: Mdust}.  These dust masses are
slightly larger than those in local galaxies with the similar stellar
mass \citep{Remy14}.  The dust masses increase with increasing stellar
mass (significance level for null hypothesis is 0.03\%).

\subsubsection{Stacking analysis of subsamples with higher/lower metallicity} \label{subsubsec: stack metal dust}
To study the metallicity dependence of dust mass, we carried out a
stacking analysis with the subsamples used in section \ref{subsubsec:
  stack metal CO}.  The stacked continuum maps for the subsamples with
lower and higher metallicity are shown in the right panel of the lower
middle and bottom rows of Figure \ref{fig: stack image}, respectively.
The dust emission is marginally detected for the subsamples with lower
metallicity, and significantly detected for the subsamples with higher
metallicity.  The derived dust masses are
$(5.6\pm2.2)\times10^{7}~M_\odot$ and
$(7.7\pm1.7)\times10^{7}~M_\odot$ for the subsamples with lower and
higher metallicity, respectively, and are plotted with blue stars in
the right panel of Figure \ref{fig: Mdust}.  Although the dust mass
seems to increase with increasing metallicity, the trend is not
significant (significance level for null hypothesis is 5.8\%).


\section{Gas-to-dust ratio} \label{sec: GDR}
\begin{figure}[t]
\begin{center}
	\epsscale{1.0}
		\plotone{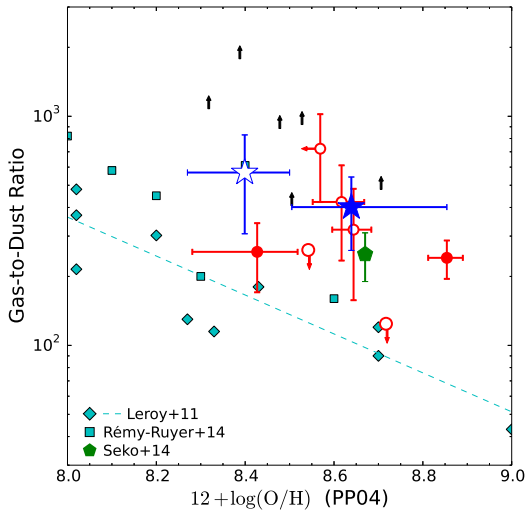}
		\caption{Gas-to-dust ratio against metallicity.
                  Filled and open red circles refer to the galaxies
                  from which dust emission is detected significantly and 
                  marginally, respectively.  
                  Black arrows show the lower limit.
                  Filled/open blue stars refer to the results of
                  stacking analysis for the subsamples with
                  higher/lower metallicity (open symbol refers to
                  marginal detection of dust).  Green pentagon refers to
                  the stacked value of three main sequence galaxies at
                  $z\sim1.4$ by \citet{Seko14}.  Cyan diamonds and
                  cyan dashed line represent local galaxies by
                  \citet{Lero11}, and cyan squares represent the
                  average values in local galaxies shown by
                  \citet{Remy14}.  (Metallicities are calibrated
                  using \citet{PP04}) }
\label{fig: GDR}
\end{center}
\end{figure}

Since the molecular gas masses and dust masses have been obtained, we
determine gas-to-dust mass ratios in the main sequence galaxies at
$z\sim1.4$, excepting seven galaxies for which only upper limits on
both the molecular gas mass and dust mass are obtained.  The derived
gas-to-dust ratios are shown in Table \ref{table: CO} and plotted in
Figure \ref{fig: GDR}.  The results of the stacking analyses for the
subsamples with lower and higher metallicity are also plotted with
blue stars.  The gas-to-dust ratios in the subsamples with lower and
higher metallicity are $568\pm261$ and $401\pm142$, respectively, and
are 3-4 times larger than those in local galaxies at a fixed
metallicity \citep{Lero11, Remy14}.  Moreover, although the
gas-to-dust ratios for local galaxies include the HI mass, the ratios
at $z\sim1.4$ do not include this, and so the gas-to-dust ratio in our
sample galaxies must be larger than the values derived here.
Therefore, care should be taken in deriving molecular gas mass from
dust mass by assuming the gas-to-dust ratio in local galaxies.

\cite{Seko14} observed CO($J=2-1$) emission lines from three main
sequence galaxies with solar metallicity at $z\sim1.4$ detected with
{\it Spitzer}/MIPS in $24~\mu\mathrm{m}$ and {\it Herschel}/SPIRE in
the $250~\mu\mathrm{m}$ and $350~\mu\mathrm{m}$, and derived
gas-to-dust ratios.  The stacked gas-to-dust ratio of the three
galaxies is $250\pm60$ (they assumed $T_\mathrm{dust}=35~\mathrm{K}$)
which is slightly larger than among local galaxies
at the same metallicity.  However, because the three galaxies observed
by \citet{Seko14} are clearly detected with SPIRE, they may be biased
to larger dust masses.  On the other hand, because almost all of our
sample galaxies in this study are detected neither with MIPS nor
SPIRE, there should be little bias to large dust mass.

It is worth mentioning that while we derive the dust mass using
a modified black body model with a fixed dust temperature and dust
emissivity index, the dust masses in the local galaxies shown in Figure
\ref{fig: GDR} are derived by adopting the models of \citet{DL07}
(hereafter DL07 models) using {Herschel}/PACS and SPIRE data.
\citet{Magd12b} showed that dust masses derived using modified black body
models with dust temperatures derived by fitting (average temperature
is $33~\mathrm{K}$) and the same fixed dust emissivity index
($\beta=1.5$) are about two times smaller than those derived by DL07
models.  Hence the gas-to-dust ratios of our samples plotted in Figure
\ref{fig: GDR} would be lower by $\sim0.3$ dex, if we used DL07 models.
The ratios are still larger by about factor of 2.  This result that
gas-to-dust ratios in the main sequence galaxies at $z\sim1.4$ are
about two times those for local galaxies is similar to
results obtained for lensed galaxies on the main sequence at
$z=1.4-3.1$ \citep{Sain13}.

The gas-to-dust ratio at $z\sim1.4$ seems to decrease with increasing
metallicity.  Although this trend is the same as that in local
galaxies \citep{Lero11, Remy14}, since the uncertainty of our results
is large, the trend is not significant 
(significance level for null hypothesis is 16\%).

\begin{figure*}[t]
\begin{center}
	\epsscale{2.0}
		\plotone{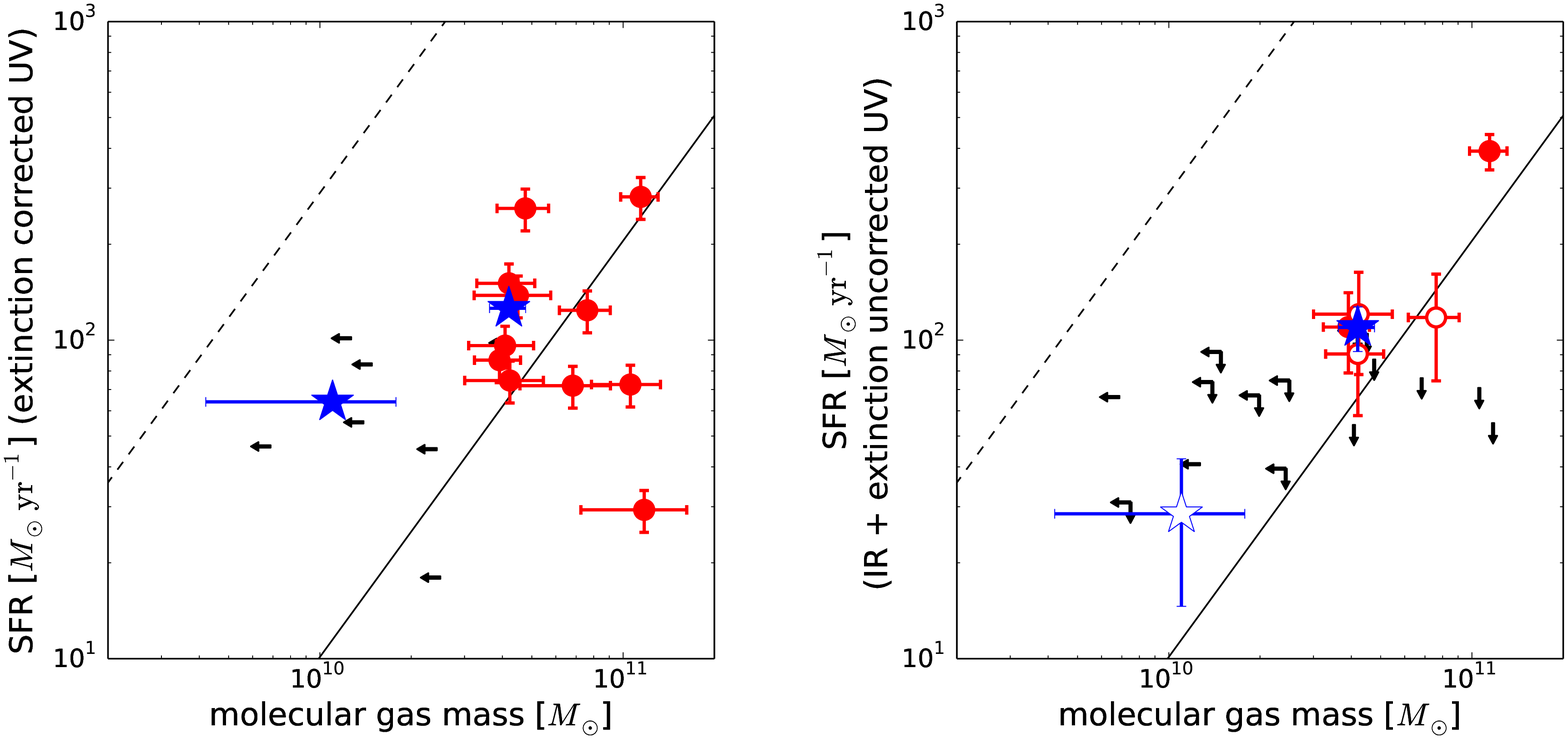}
		\caption{SFR against molecular gas mass. 
		(Left) SFRs are derived from extinction corrected UV luminosity densities. 
		Filled (red) circles refer to the CO detected galaxies 
		and (black) left arrows show the CO non-detected galaxies. 
		Filled (blue) stars refer to the results of stacking analysis for the subsamples 
		of CO detected and non-detected. 
		(Right) SFRs are sum of the SFRs from total infrared luminosities and the 
		SFRs from extinction uncorrected UV luminosity densities. 
		Filled (red) circles refer to the both CO and dust detected galaxies. 
		Down-arrows show the dust non-detected galaxies but CO emissions are detected. 
		Stars and left arrows are the same as those in the left panel. 
		Open symbols represent the marginal detections of dust continuum emission. 
		Solid and dashed lines represent the sequences of normal star-forming 
		galaxies (e.g., local spirals and sBzK galaxies)
		and starburst galaxies (e.g., local ULIRGs and SMGs), respectively, 
		given by \citet{Dadd10b}. 
		}
\label{fig: KS-law}
\end{center}
\end{figure*}

\section{Molecular gas mass against SFR, and gas depletion time}
We examine the location of our sample galaxies 
in the molecular gas mass$-$SFR diagram. 
In the left panel of Figure~\ref{fig: KS-law}, 
the SFRs are derived from extinction corrected UV luminosity densities. 
In the right panel of Figure~\ref{fig: KS-law}, 
the SFRs are sum of the SFRs from total infrared luminosities 
\citep[$L_\mathrm{IR} (8-1000~\mu\mathrm{m})$;][]{Kenn98}
and the SFRs from extinction uncorrected UV luminosity densities, 
and are listed in Table~\ref{table: sample}. 
$L_\mathrm{IR}$ are derived by fitting a template SED 
of main sequence galaxies at $z\sim1.5$ \citep{Magd12b} 
to the continuum data obtained in this observation. 
For the dust detected galaxies, the SFRs from $L_\mathrm{IR}$ 
and extinction uncorrected UV luminosity densities roughly 
agree with those from extinction corrected UV luminosity densities. 
In this figure, the solid line represents the sequence of 
normal star-forming galaxies (e.g., local spiral galaxies and sBzK galaxies), 
and the dashed line represents the sequence for starburst galaxies 
(e.g., local ultra luminous infrared galaxies (ULIRGs) 
and distant submillimeter galaxies (SMGs)) given by \citet{Dadd10b}. 
Most of CO detected galaxies are located around the solid line. 

We carried out a stacking analysis of CO emissions for 
subsamples with CO detected galaxies ($M_\mathrm{mol}\geq3.9\times10^{10}~M_\odot$) 
and CO non-detected galaxies (almost all of them are $M_\mathrm{mol}<2.5\times10^{10}~M_\odot$). 
We do not include SXDS1\_35572 and SXDS1\_79307 as discussed previously. 
The CO emissions are significantly detected for both subsamples. 
The resulting molecular gas masses of CO detected 
and non-detected subsamples are $(4.2\pm0.6)\times10^{10}~M_\odot$
and $(1.1\pm0.7)\times10^{10}~M_\odot$, respectively. 
The results of stacking analysis are plotted 
in the left panel of Figure~\ref{fig: KS-law} (filled blue stars).
The SFRs derived from extinction corrected UV luminosity densities 
are taken to be average values of the stacked galaxies. 
The results of the stacking analysis for the subsamples 
of CO detected and CO non-detected galaxies are located slightly 
above the sequence of normal star-forming galaxies 
and at the middle of the sequences 
of normal star-forming galaxies and starburst galaxies, respectively. 

To derive the SFRs from the $L_\mathrm{IR}$, 
we also carried out a stacking analysis for the same subsamples 
to see the continuum emission. 
The dust emission is significantly detected for the subsamples 
with CO detected, and marginally detected for the subsamples 
with CO non-detected. 
The resulting SFRs of CO detected and non-detected 
subsamples are $(109\pm18)~M_\odot~\mathrm{yr^{-1}}$ 
and $(28\pm14)~M_\odot~\mathrm{yr^{-1}}$, respectively. 
Here, the SFRs derived from extinction uncorrected UV luminosity densities 
are taken to be average values of the stacked galaxies.
The results of stacking analysis are plotted 
in the right panel of Figure~\ref{fig: KS-law} (blue stars), 
and are located slightly above the sequence 
of normal star-forming galaxies.

\begin{figure*}
\begin{center}
	\epsscale{2.0}
		\plotone{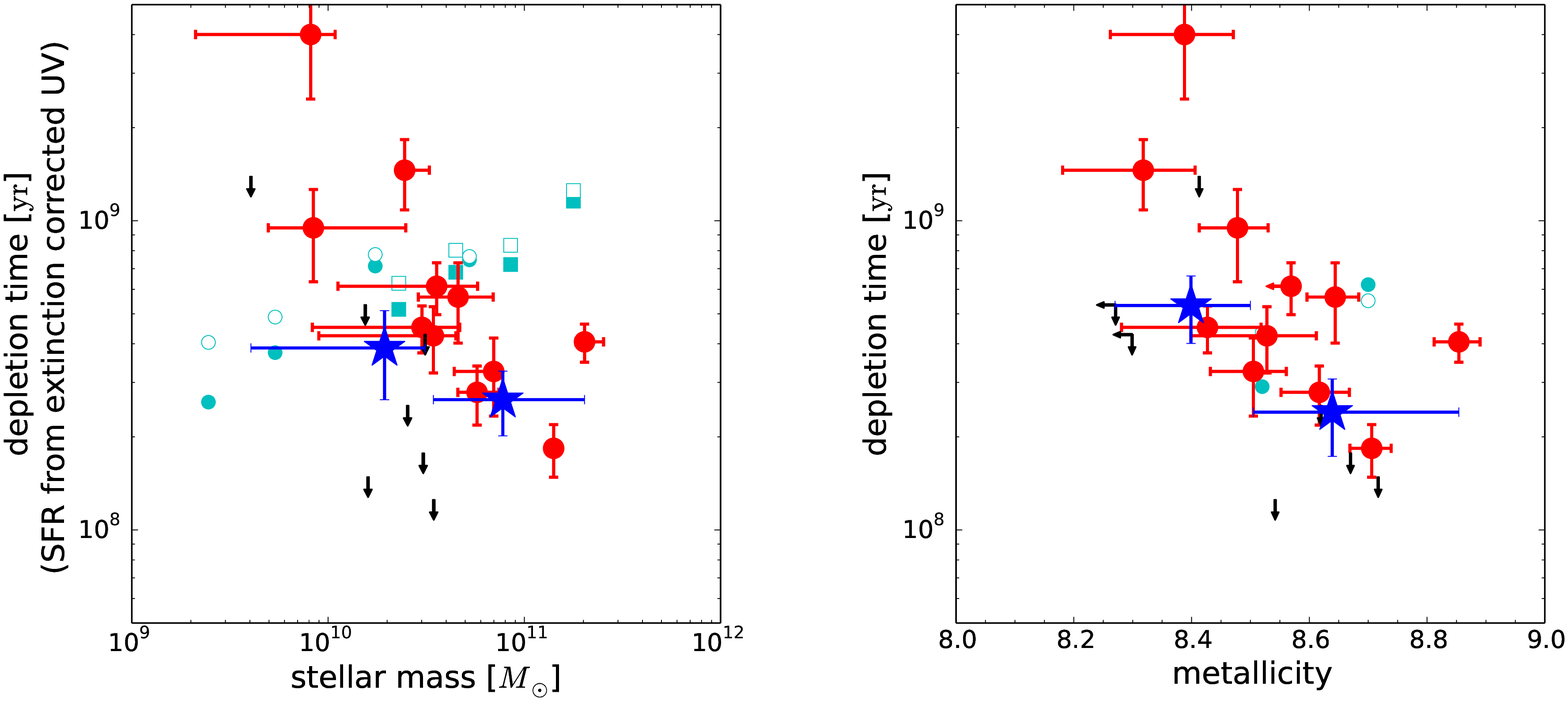}
		\plotone{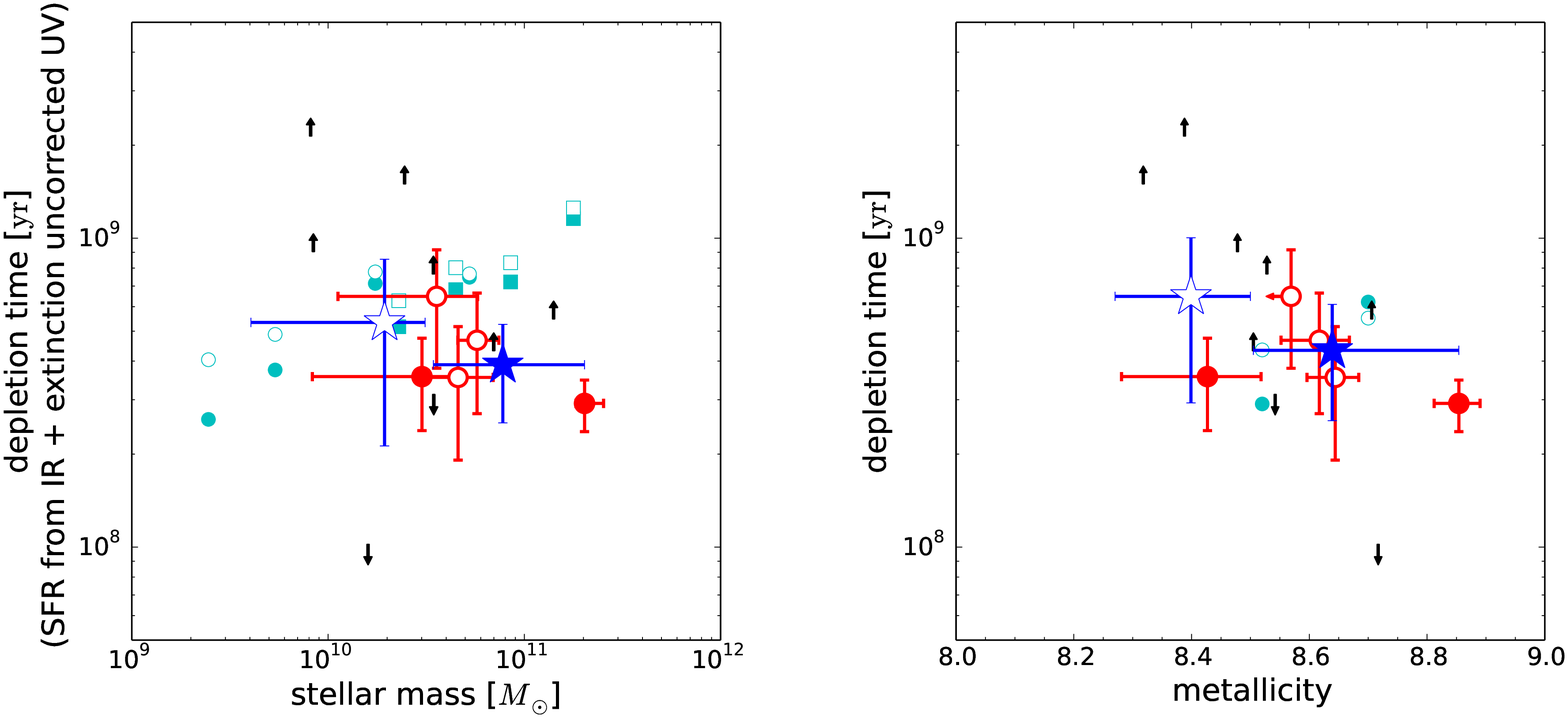}
		\caption{Depletion time of molecular gas calculated with SFR 
		from extinction corrected UV luminosity density against stellar 
		mass (upper left) and against metallicity (upper right). 
		Symbols are the same as those in Figure \ref{fig: Mgas}. 
		Depletion time of molecular gas calculated with SFR from 
		total infrared luminosity and extinction uncorrected UV luminosity 
		density against stellar mass (lower left) and against metallicity (lower right). 
		Filled (red) circles refer to the both CO and dust detected galaxies. 
		Open (red) circles refer to the CO detected and dust marginally detected galaxies. 
		Filled/open blue stars refer to the results of stacking analysis (open symbol refers to
		galaxies with marginally detected dust emission).
		Arrows show the upper and lower limits.}
\label{fig: tdep}
\end{center}
\end{figure*}

The depletion time of molecular gas is derived by 
\begin{equation}
t_\mathrm{dep} = \frac{M_\mathrm{mol}}{\mathrm{SFR}}. 
\end{equation}
The derived depletion times are plotted against stellar mass 
and metallicity in upper panels of Figure~\ref{fig: tdep}. 
Here, SFRs are derived from extinction corrected UV luminosity densities. 
For the detected galaxies, the depletion time decreases 
with increasing stellar mass and metallicity. 

We carried out a stacking analysis for subsamples with smaller and
larger stellar mass which were the same subsamples in section
\ref{subsubsec: stack mass CO}.  The SFR is taken to be the average of
the values of the stacked galaxies.  The resulting depletion times are
$(3.9\pm1.2)\times10^{8}~\mathrm{yr}$ and
$(2.6\pm0.6)\times10^{8}~\mathrm{yr}$ for the smaller and larger
stellar mass subsamples, respectively, and are plotted against stellar
mass in upper left panel of Figure~\ref{fig: tdep} (filled blue stars).
These values are significantly smaller than those in local
star-forming galaxies with similar stellar mass
\citep[e.g.,][]{Sain11b, Bose14b}. 
While the depletion time increases with stellar mass in local galaxies, 
it seems to decrease in the $z\sim1.4$ star-forming galaxies, 
though the trend is not so significant
(significance level for null hypothesis is 1.0\%). 

We carried out a stacking analysis for subsamples with lower 
and higher metallicity which were the same subsamples in
section \ref{subsubsec: stack metal CO}.  The resulting depletion times
are $(5.3\pm1.3)\times10^{8}~\mathrm{yr}$ and
$(2.4\pm0.7)\times10^{8}~\mathrm{yr}$ for the lower and higher
metallicity subsamples, respectively, and are plotted against
metallicity in upper right panel of Figure~\ref{fig: tdep} (filled blue
stars).  The depletion time decreases with increasing metallicity
(significance level for null hypothesis is 0.07\%),
while it seems to increase in local star-forming galaxies.

In lower panels of Figure~\ref{fig: tdep}, we also plot depletion times 
calculated with SFRs from $L_\mathrm{IR}$ and extinction 
uncorrected UV luminosity densities. 
From the stacking analysis, SFRs are $(45\pm12)~M_\odot~\mathrm{yr^{-1}}$ 
and $(105\pm21)~M_\odot~\mathrm{yr^{-1}}$ for the smaller 
and larger stellar mass subsamples, respectively, 
and the resulting depletion times are $(5.3\pm3.2)\times10^{8}~\mathrm{yr}$ and 
$(3.9\pm1.4)\times10^{8}~\mathrm{yr}$, respectively (lower left panel of 
Figure~\ref{fig: tdep} (blue stars)). 
The depletion time seems to decrease with increasing stellar mass, 
though the trend is not significant
(significance level for null hypothesis is 24\%). 
SFRs are $(50\pm16)~M_\odot~\mathrm{yr^{-1}}$ 
and $(72\pm13)~M_\odot~\mathrm{yr^{-1}}$ for the lower 
and higher metallicity subsamples, respectively, 
and the resulting depletion times are $(6.4\pm3.6)\times10^{8}~\mathrm{yr}$ and 
$(4.3\pm1.8)\times10^{8}~\mathrm{yr}$, respectively, (lower right panel of 
Figure~\ref{fig: tdep} (blue stars)). 
Although the depletion time seems to decrease with increasing metallicity, 
the trend is not significant
(significance level for null hypothesis is 19\%).
The similar trend is also seen. 

In all cases, the depletion time tends to decrease with increasing stellar mass
and metallicity, which contrasts with the trends in local star-forming galaxies; 
the depletion time increases with increasing stellar mass and metallicity
\citep[e.g.,][]{Sain11b, Bose14b}. 
It means that the average star formation efficiency 
of galaxies with larger stellar mass and metallicity 
is higher at high redshift, but lower in local universe 
compared with that of galaxies with smaller 
stellar mass and metallicity.

\section{Summary}
To investigate the properties of the interstellar medium in
star-forming galaxies at the noon of star-formation rate density, we
carried out observations of $^{12}\mathrm{CO}(J=5-4)$ and dust thermal
continuum emission toward 20 main sequence galaxies at $z\sim1.4$
using ALMA.  Gas-phase metallicities in these galaxies were derived from
near-infrared spectroscopic observations with FMOS on the Subaru
telescope using H$\alpha$ and [NII] emission lines.  The range of
stellar masses (adopting Salpeter IMF) and metallicities
($12+\log(\mathrm{O/H})$) of our sample are $4\times10^{9} -
4\times10^{11}~M_\odot$ and $8.2-8.9$, respectively, and these
galaxies uniformly trace the main sequence and mass$-$metallicity
relation at this redshift.  The stellar mass range covers a lower mass
than that reported in previous studies.

We detected CO emission lines from 11 galaxies.  The CO lines tend to
be detected for galaxies with more massive/higher SFRs.  No clear
dependence on metallicity is seen, although the average metallicity of
the detected galaxies is slightly larger than that of non-detected
galaxies.  Molecular gas masses are derived by assuming the
CO($5-4$)/CO($1-0$) luminosity ratio of 0.23, and by adopting a
metallicity-dependent CO-to-H$_2$ conversion factor.  Molecular gas
masses and their fraction against stellar mass for the detected
galaxies range from $(3.9-12)\times10^{10}~M_\odot$ and $0.25-0.94$,
respectively; these values are significantly larger than those in
local spiral galaxies.  Since the CO emission lines from about half of
our sample galaxies are not detected, we performed stacking analyses
to examine the relations of the molecular gas mass and its fraction
against stellar mass and metallicity.  The results of stacking
analyses using the whole sample of galaxies are that the molecular gas
mass is $(2-4)\times10^{10}~M_\odot$ and increases with increasing
stellar mass, but does not depend on metallicity.  The molecular gas
mass fraction is $30-65\%$ and decreases both with increasing stellar
mass and metallicity.  Due to the mass$-$metallicity relation, the
dependencies on stellar mass and metallicity are not separated
clearly.  To avoid the metallicity effect, we performed stacking
analyses for subsamples with smaller and larger stellar mass at the
almost the same metallicity.  The molecular gas mass increases with
increasing stellar mass.  Its fraction seems to decrease with
increasing stellar mass, but the trend is not so significant.  To
avoid the stellar mass effect, we made stacking analyses for
subsamples with lower and higher metallicity at the almost same
stellar mass.  Both the molecular gas mass and its fraction decrease
with increasing metallicity.  

We detected significant continuum emissions from 2 galaxies with
marginal detections from a further 5 galaxies.  Continuum emissions
tend to be detected for galaxies with larger stellar mass and
with higher metallicity.  Dust masses were derived with a modified black
body model by adopting a dust temperature of $30~\mathrm{K}$ and a
dust emissivity index of $1.5$.  The derived dust masses of the
detected galaxies are $(3.9-38)\times10^{7}~M_\odot$.  We used 
stacking analyses to examine the relations of the dust mass against
stellar mass and metallicity.  The dust mass increases with increasing stellar
mass and seems to increase with increasing metallicity.

We also derive the gas-to-dust ratio.  The result of stacking 
the subsamples with lower and higher metallicity shows that the
gas-to-dust ratios are $\sim500$, which is 3-4 times larger than those
in local galaxies at the same metallicity range.  The gas-to-dust
ratio at $z\sim1.4$ seems to decrease with increasing metallicity.

CO detected galaxies locate around the sequence of normal star-forming 
galaxies in the molecular gas mass$-$SFR diagram. 
The results of stacking analysis show that the depletion time of molecular gas is 
$\sim3\times10^{8}~\mathrm{yr}$ which is smaller than that in local galaxies. 
Although the uncertainty of depletion time is large, 
the timescale tends to decrease with increasing stellar mass and metallicity.
These trends contrast with those in local star-forming galaxies.

\acknowledgments
We would like to thank the referee for useful comments and suggestions.
We are grateful to Kazuya Saigo and the staff at the ALMA Regional Center for their help in data reduction.
A.S. is supported by Research Fellowship for Young Scientists from the Japan Society of the Promotion of Science (JSPS). 
K.O. was supported by Grant-in-Aid for Scientific Research (C) (24540230) from JSPS.
Kavli IPMU is supported by World Premier International Research Center Initiative (WPI), MEXT, Japan. 
This paper makes use of the following ALMA data: ADS/JAO.ALMA\#2011.0.00648.S. 
ALMA is a partnership of ESO (representing its member states), NSF (USA), and NINS (Japan), together with NRC (Canada) and NSC and ASIAA (Taiwan), 
in cooperation with the Republic of Chile. The Joint ALMA Observatory is operated by ESO, AUI/NRAO, and NAOJ.

{\it Facilities:} \facility{ALMA}.

%
%

\clearpage

\begin{deluxetable}{lcccccccc}
\tabletypesize{\scriptsize}
\tablecaption{Sample galaxies.}
\tablewidth{0pt}
\tablehead{
\colhead{ID} & \colhead{RA\tablenotemark{a}} & \colhead{DEC\tablenotemark{a}} & \colhead{$z_\mathrm{spec}$\tablenotemark{b}} & \colhead{$M_\ast$\tablenotemark{c}} &
\colhead{metallicity\tablenotemark{d}} & \colhead{SFR\tablenotemark{c}} & \colhead{SFR\tablenotemark{e}} & \colhead{SFR\tablenotemark{f}}\\
\colhead{} & \colhead{(J2000)} & \colhead{(J2000)} & \colhead{} & \colhead{($M_\odot$)} &
\colhead{$12+\log \mathrm{(O/H)}$} & \colhead{($M_\odot~\mathrm{yr^{-1}}$)} & \colhead{($M_\odot~\mathrm{yr^{-1}}$)} & \colhead{($M_\odot~\mathrm{yr^{-1}}$)}}
\startdata
SXDS1\_13015		& 02:17:13.63 & $-$05:09:39.8 & 1.451 & $(2.0_{-0.2}^{+0.5})\times10^{11}$	& $8.85_{-0.04}^{+0.04}$	& 282	& 154 & $392\pm50$ \\
SXDS1\_1723		& 02:17:32.70 & $-$05:13:16.5 & 1.467 & $(3.1_{-2.0}^{+3.0})\times10^{10}$	& $<8.30$				& 98		& 154 & $< 107$ \\
SXDS1\_31189		& 02:17:13.68 & $-$05:04:07.7 & 1.394 & $(8.2_{-6.0}^{+2.7})\times10^{9}$	& $8.39_{-0.13}^{+0.08}$	& 29		& 72 & $< 55$ \\
SXDS1\_33244		& 02:16:47.40 & $-$05:03:28.1 & 1.474 & $(5.7_{-1.2}^{+1.7})\times10^{10}$	& $8.62_{-0.07}^{+0.05}$	& 151	& 189 & $90\pm33$ \\
SXDS1\_35572		& 02:17:34.65 & $-$05:02:39.0 & 1.347 & $(3.8_{-1.1}^{+0.1})\times10^{11}$	& $8.67_{-0.09}^{+0.08}$	& 4938	& 537 & $< 31$ \\
SXDS1\_42087		& 02:17:24.36 & $-$05:00:44.9 & 1.594 & $(3.6_{-2.5}^{+2.2})\times10^{10}$	& $<8.57$				& 124	& 143 & $118\pm43$ \\
SXDS1\_59863		& 02:17:45.88 & $-$04:54:37.6 & 1.448 & $(1.4_{-0.1}^{+0.1})\times10^{11}$	& $8.71_{-0.04}^{+0.03}$	& 259	& 228 & $< 87$ \\
SXDS1\_59914		& 02:17:12.98 & $-$04:54:40.4 & 1.460 & $(7.0_{-2.6}^{+0.0})\times10^{10}$	& $8.51_{-0.07}^{+0.06}$	& 138	& 188 & $< 105$ \\
SXDS1\_67002		& 02:19:02.65 & $-$04:49:55.9 & 1.281 & $(3.5_{-0.1}^{+4.7})\times10^{10}$	& $8.54_{-0.08}^{+0.06}$	& 101	& 88 & $41\pm13$ \\
SXDS1\_68849		& 02:17:00.28 & $-$04:48:14.5 & 1.325 & $(2.5_{-1.9}^{+2.1})\times10^{10}$	& $8.62_{-0.05}^{+0.04}$	& 55		& 70 & $< 74$ \\
SXDS1\_79307		& 02:17:05.79 & $-$04:51:25.7 & 1.575 & $(2.1_{-0.0}^{+0.0})\times10^{11}$	& $<8.38$				& 3822	& 10474 & $< 67$ \\
SXDS1\_79518		& 02:18:59.06 & $-$04:51:24.9 & 1.330 & $(2.5_{-0.2}^{+0.8})\times10^{10}$	& $8.32_{-0.14}^{+0.09}$	& 73		& 146 & $< 71$ \\
SXDS2\_13316		& 02:17:39.03 & $-$04:44:41.8 & 1.446 & $(8.4_{-3.5}^{+16})\times10^{9}$	& $8.48_{-0.07}^{+0.05}$	& 72		& 118 & $< 76$ \\
SXDS2\_22198		& 02:17:53.42 & $-$04:42:53.4 & 1.499 & $(1.6_{-0.1}^{+0.1})\times10^{10}$	& $8.72_{-0.09}^{+0.07}$	& 46		& 87 & $66\pm17$ \\
SXDS3\_101746	& 02:18:04.18 & $-$05:19:38.3 & 1.335 & $(4.0_{-0.8}^{+1.3})\times10^{9}$	& $8.41_{-0.09}^{+0.07}$	& 18		& 70 & $< 75$ \\
SXDS3\_103139	& 02:16:57.65 & $-$05:14:34.9 & 1.382 & $(1.5_{-0.1}^{+0.7})\times10^{10}$	& $<8.27$				& 45		& 68 & $< 39$ \\
SXDS3\_110465	& 02:18:20.95 & $-$05:19:07.7 & 1.458 & $(3.1_{-0.8}^{+2.3})\times10^{10}$	& $8.67_{-0.03}^{+0.03}$	& 84		& 90 & $< 92$ \\
SXDS5\_19723		& 02:16:24.37 & $-$05:09:18.1 & 1.533 & $(3.4_{-2.5}^{+1.0})\times10^{10}$	& $8.53_{-0.13}^{+0.08}$	& 96		& 162 & $< 54$ \\
SXDS5\_28019		& 02:16:08.53 & $-$05:06:15.6 & 1.348 & $(3.0_{-2.2}^{+1.7})\times10^{10}$	& $8.43_{-0.15}^{+0.09}$	& 87		& 104 & $110\pm31$ \\
SXDS5\_9364		& 02:16:33.81 & $-$05:13:44.7 & 1.441 & $(4.6_{-1.7}^{+2.3})\times10^{10}$	& $8.64_{-0.05}^{+0.04}$	& 75		& 102 & $121\pm43$ \\
\enddata
\label{table: sample}
\tablenotetext{a}{The accuracy of coordinates in the K-band image is $\sim0''.2-0''.3$. }
\tablenotetext{b}{Derived from H$\alpha$ wavelength in vacuum. The error is typically $\pm0.001$. }
\tablenotetext{c}{We adopted Salpeter IMF \citep{Salp55}. 
			SFR is derived from extinction corrected UV luminosity density. 
			The error of SFR is typically $15\%$. }
\tablenotetext{d}{Derived from H$\alpha$ and [NII]$\lambda~6584$ calibrated by \citet{PP04}. }
\tablenotetext{e}{SFR is derived from extinction corrected H$\alpha$ luminosity density. 
			The error of SFR is typically $10\%$. }
\tablenotetext{f}{Sum of SFR from $L_\mathrm{IR} (8-1000~\mu\mathrm{m})$ 
			from our dust continuum observations 
			and SFR from extinction uncorrected UV luminosity densities.}
\end{deluxetable}

\begin{deluxetable}{lccccccc}
\tabletypesize{\scriptsize}
\tablecaption{Results of the observations.}
\tablewidth{0pt}
\tablehead{
\colhead{ID}				& \colhead{$\int S_\mathrm{CO(5-4)} dv$}	& \colhead{$L_\mathrm{CO(5-4)}^{'}$}	& \colhead{$M_\mathrm{gas}$\tablenotemark{a}}		
& \colhead{$f_\mathrm{gas}$\tablenotemark{b}}	& \colhead{$S_\mathrm{continuum}$\tablenotemark{c}}		&\colhead{$M_\mathrm{dust}$\tablenotemark{d}}			& Gas-to-Dust Ratio	\\
\colhead{}					& \colhead{($\mathrm{Jy~km~s^{-1}}$)}		& \colhead{($10^{9}~\mathrm{K~km~s^{-1}~pc^{2}}$)}	& \colhead{($10^{10}~M_\odot$)}	
& \colhead{}				& \colhead{(mJy)}						&\colhead{($10^{7}~M_\odot$)}			& \colhead{}	}
\startdata
SXDS1\_13015		& $2.31\pm0.33$	& $10\pm1$	& $11\pm2$	& $0.36\pm0.05$	& $0.86\pm0.11$	& $47\pm6$	& $241\pm46$	\\
SXDS1\_1723		& $<0.20$			& $<0.9$		& $<4.2$		& $<0.57$			& $<0.20$			& $<11$		& -----	\\
SXDS1\_31189		& $0.86\pm0.33$	& $3.5\pm1.3$	& $12\pm4$	& $0.94\pm0.04$	& $<0.12$			& $<6.6$		& $>1783$	\\
SXDS1\_33244		& $0.52\pm0.11$	& $2.4\pm0.5$	& $4.2\pm0.9$	& $0.42\pm0.08$	& $0.18\pm0.07$	& $9.9\pm3.9$	& $423\pm188$\\
SXDS1\_35572		& $<0.12$			& $<0.5$		& $<0.8$		& $<0.02$			& $<0.08$			& $<4.4$		& -----	\\
SXDS1\_42087		& $0.73\pm0.14$	& $3.8\pm0.7$	& $7.6\pm1.5$	& $0.68\pm0.15$	& $0.19\pm0.07$	& $11\pm4$	& $722\pm300$\\
SXDS1\_59863		& $0.74\pm0.14$	& $3.2\pm0.6$	& $4.8\pm0.9$	& $0.25\pm0.04$	& $<0.18$			& $<9.9$		& $>480$		\\
SXDS1\_59914		& $0.43\pm0.12$	& $1.9\pm0.5$	& $4.5\pm1.3$	& $0.39\pm0.08$	& $<0.20$			& $<11$		& $>408$		\\
SXDS1\_67002		& $<0.17$			& $<0.6$		& $<1.3$		& $<0.27$			& $0.09\pm0.04$	& $4.9\pm2.2$	& $<261$		\\
SXDS1\_68849		& $<0.21$			& $<0.8$		& $<1.4$		& $<0.35$			& $<0.18$			& $<9.8$		& -----		\\
SXDS1\_79307		& $<0.11$			& $<0.6$		& $<2.0$		& $<0.09$			& $<0.12$			& $<6.7$		& -----		\\
SXDS1\_79518		& $0.66\pm0.17$	& $2.4\pm0.6$	& $11\pm3$	& $0.81\pm0.05$	& $<0.18$			& $<9.8$		& $>1079$	\\
SXDS2\_13316		& $0.62\pm0.21$	& $2.7\pm0.9$	& $6.8\pm2.3$	& $0.89\pm0.12$	& $<0.14$			& $<7.7$		& $>885$		\\
SXDS2\_22198		& $<0.10$			& $<0.5$		& $<0.7$		& $<0.30$			& $0.10\pm0.04$	& $5.5\pm2.2$	& $<124$		\\
SXDS3\_101746	& $<0.22$			& $<0.8$		& $<2.5$		& $<0.86$			& $<0.18$			& $<9.8$		& -----		\\
SXDS3\_103139	& $<0.12$			& $<0.5$		& $<2.4$		& $<0.61$			& $<0.08$			& $<4.4$		& -----		\\
SXDS3\_110465	& $<0.21$			& $<0.9$		& $<1.5$		& $<0.33$			& $<0.18$			& $<9.9$		& -----		\\
SXDS5\_19723		& $0.38\pm0.09$	& $1.9\pm0.4$	& $4.1\pm1.0$	& $0.54\pm0.14$	& $<0.08$			& $<4.4$		& $>921$		\\
SXDS5\_28019		& $0.35\pm0.06$	& $1.3\pm0.2$	& $3.9\pm0.7$	& $0.57\pm0.16$	& $0.28\pm0.08$	& $15\pm4$	& $256\pm86$	\\
SXDS5\_9364		& $0.58\pm0.17$	& $2.5\pm0.7$	& $4.2\pm1.2$	& $0.48\pm0.13$	& $0.24\pm0.10$	& $13\pm6$	& $320\pm162$\\
\enddata
\label{table: CO}
\tablenotetext{a}{We adopted the CO(5-4)/CO(1-0) luminosity ratio of 0.23 \citep{Dadd15} and metallicity-dependent CO-to-H$_2$ conversion factor shown with the equation (\ref{eq: conversion factor}) \citep{Genz12}. }
\tablenotetext{b}{$f_\mathrm{gas} = \frac{M_\mathrm{gas}}{M_\mathrm{gas} + M_\mathrm{star}}$}
\tablenotetext{c}{The average observed wavelength is $\sim1.3~\mathrm{mm}$, thus, the average rest-frame wavelength is $\sim0.5~\mathrm{mm}$.}
\tablenotetext{d}{We used a modified black body model adopting a dust temperature of $30~\mathrm{K}$ and a dust emissivity index of 1.5. }
\end{deluxetable}

\begin{deluxetable}{lcc}
\tabletypesize{\scriptsize}
\tablecaption{Results of the observations. (continued)}
\tablewidth{0pt}
\tablehead{
\colhead{ID}				& \colhead{noise level\tablenotemark{e}}				&\colhead{beam size, PA}\\
\colhead{}					& \colhead{(mJy/beam)}			&\colhead{}}
\startdata
SXDS1\_13015		& 0.09	& $0''.85\times0''.65$, $108^{\circ}$	\\
SXDS1\_1723		& 0.10	& $1''.28\times0''.64$, $72^{\circ}$	\\
SXDS1\_31189		& 0.06	& $0''.80\times0''.66$, $100^{\circ}$	\\
SXDS1\_33244		& 0.07	& $1''.04\times0''.64$, $75^{\circ}$	\\
SXDS1\_35572		& 0.04	& $0''.84\times0''.68$, $93^{\circ}$	\\
SXDS1\_42087		& 0.06	& $0''.80\times0''.66$, $96^{\circ}$	\\
SXDS1\_59863		& 0.09	& $0''.85\times0''.65$, $110^{\circ}$	\\
SXDS1\_59914		& 0.10	& $1''.20\times0''.65$, $73^{\circ}$	\\
SXDS1\_67002		& 0.07	& $0''.87\times0''.63$, $114^{\circ}$	\\
SXDS1\_68849		& 0.09	& $0''.96\times0''.65$, $76^{\circ}$	\\
SXDS1\_79307		& 0.06	& $0''.80\times0''.66$, $93^{\circ}$	\\
SXDS1\_79518		& 0.09	& $1''.02\times0''.65$, $75^{\circ}$	\\
SXDS2\_13316		& 0.07	& $0''.86\times0''.64$, $109^{\circ}$	\\
SXDS2\_22198		& 0.04	& $0''.86\times0''.68$, $86^{\circ}$	\\
SXDS3\_101746	& 0.09	& $1''.06\times0''.65$, $74^{\circ}$	\\
SXDS3\_103139	& 0.04	& $0''.83\times0''.68$, $100^{\circ}$	\\
SXDS3\_110465	& 0.09	& $1''.13\times0''.65$, $74^{\circ}$	\\
SXDS5\_19723		& 0.04	& $0''.83\times0''.68$, $97^{\circ}$	\\
SXDS5\_28019		& 0.04	& $0''.85\times0''.68$, $90^{\circ}$	\\
SXDS5\_9364		& 0.08	& $0''.87\times0''.64$, $112^{\circ}$	\\
\enddata
\label{table: CO-2}
\tablenotetext{e}{Noise level in continuum map. }
\end{deluxetable}

\end{document}